\documentclass[prb,twocolumn,showpacs,preprintnumbers,amsmath,aps]{revtex4}
\usepackage{graphicx,hyperref}% Include figure files
\usepackage{bm}% bold math
\usepackage{subfigure}% bold math

\newcommand\dlc{d_{\text{lc}}}

\newcommand\ie{{\it i.e.}}
\newcommand\eg{{\it e.g.}}

\newcommand\etab{\bar \eta}

\newcommand\vect[1]{ \mathbf{ #1}}
\def\nbC{{\mathchoice {\setbox0=\hbox{$\displaystyle\rm C$}%
\hbox{\hbox to0pt{\kern0.4\wd0\vrule height0.9\ht0\hss}\box0}}
{\setbox0=\hbox{$\textstyle\rm C$}\hbox{\hbox
to0pt{\kern0.4\wd0\vrule height0.9\ht0\hss}\box0}}
{\setbox0=\hbox{$\scriptstyle\rm C$}\hbox{\hbox
to0pt{\kern0.4\wd0\vrule height0.9\ht0\hss}\box0}}
{\setbox0=\hbox{$\scriptscriptstyle\rm C$}\hbox{\hbox
to0pt{\kern0.4\wd0\vrule height0.9\ht0\hss}\box0}}}}
\def\nbQ{{\mathchoice {\setbox0=\hbox{$\displaystyle\rm
Q$}\hbox{\raise
0.15\ht0\hbox to0pt{\kern0.4\wd0\vrule height0.8\ht0\hss}\box0}}
{\setbox0=\hbox{$\textstyle\rm Q$}\hbox{\raise
0.15\ht0\hbox to0pt{\kern0.4\wd0\vrule height0.8\ht0\hss}\box0}}
{\setbox0=\hbox{$\scriptstyle\rm Q$}\hbox{\raise
0.15\ht0\hbox to0pt{\kern0.4\wd0\vrule height0.7\ht0\hss}\box0}}
{\setbox0=\hbox{$\scriptscriptstyle\rm Q$}\hbox{\raise
0.15\ht0\hbox to0pt{\kern0.4\wd0\vrule height0.7\ht0\hss}\box0}}}}
\def\nbT{{\mathchoice {\setbox0=\hbox{$\displaystyle\rm
T$}\hbox{\hbox to0pt{\kern0.3\wd0\vrule height0.9\ht0\hss}\box0}}
{\setbox0=\hbox{$\textstyle\rm T$}\hbox{\hbox
to0pt{\kern0.3\wd0\vrule height0.9\ht0\hss}\box0}}
{\setbox0=\hbox{$\scriptstyle\rm T$}\hbox{\hbox
to0pt{\kern0.3\wd0\vrule height0.9\ht0\hss}\box0}}
{\setbox0=\hbox{$\scriptscriptstyle\rm T$}\hbox{\hbox
to0pt{\kern0.3\wd0\vrule height0.9\ht0\hss}\box0}}}}
\def\nbS{{\mathchoice
{\setbox0=\hbox{$\displaystyle     \rm S$}\hbox{\raise0.5\ht0%
\hbox to0pt{\kern0.35\wd0\vrule height0.45\ht0\hss}\hbox
to0pt{\kern0.55\wd0\vrule height0.5\ht0\hss}\box0}}
{\setbox0=\hbox{$\textstyle        \rm S$}\hbox{\raise0.5\ht0%
\hbox to0pt{\kern0.35\wd0\vrule height0.45\ht0\hss}\hbox
to0pt{\kern0.55\wd0\vrule height0.5\ht0\hss}\box0}}
{\setbox0=\hbox{$\scriptstyle      \rm S$}\hbox{\raise0.5\ht0%
\hboxto0pt{\kern0.35\wd0\vrule height0.45\ht0\hss}\raise0.05\ht0%
\hbox to0pt{\kern0.5\wd0\vrule height0.45\ht0\hss}\box0}}
{\setbox0=\hbox{$\scriptscriptstyle\rm S$}\hbox{\raise0.5\ht0%
\hboxto0pt{\kern0.4\wd0\vrule height0.45\ht0\hss}\raise0.05\ht0%
\hbox to0pt{\kern0.55\wd0\vrule height0.45\ht0\hss}\box0}}}}
\def\nbZ{{\mathchoice {\hbox{$\sf\textstyle Z\kern-0.4em Z$}}
{\hbox{$\sf\textstyle Z\kern-0.4em Z$}}
{\hbox{$\sf\scriptstyle Z\kern-0.3em Z$}}
{\hbox{$\sf\scriptscriptstyle Z\kern-0.2em Z$}}}}
\begin{document}

\title{Two-loop Functional Renormalization Group of the Random Field
  and Random Anisotropy $O(N)$ Models}

\author{Matthieu Tissier} \email{tissier@lptl.jussieu.fr}
\affiliation{LPTMC, CNRS-UMR 7600, Universit\'e Pierre et Marie Curie,
bo\^ite 121, 4 Pl. Jussieu, 75252 Paris c\'edex 05, France}

\author{Gilles Tarjus} \email{tarjus@lptl.jussieu.fr}
\affiliation{LPTMC, CNRS-UMR 7600, Universit\'e Pierre et Marie Curie,
bo\^ite 121, 4 Pl. Jussieu, 75252 Paris c\'edex 05, France}

\date{\today}

\begin{abstract}
  We study by the perturbative Functional Renormalization Group (FRG)
  the Random Field and Random Anisotropy $O(N)$ models near $d=4$, the
  lower critical dimension of ferromagnetism. The long-distance
  physics is controlled by zero-temperature fixed points at which the
  renormalized effective action is nonanalytic. We obtain the beta
  functions at 2-loop order, showing that despite the nonanalytic
  character of the renormalized effective action, the theory is
  perturbatively renormalizable at this order. The physical results
  obtained at 2-loop level, most notably concerning the breakdown of
  dimensional reduction at the critical point and the stability of
  quasi-long range order in $d<4$, are shown to fit into the picture
  predicted by our recent nonperturbative FRG approach.

\end{abstract}


\maketitle

\section{Introduction}\label{sec:introduction}

Despite decades of intensive investigation the effect of weak quenched
disorder on the long-distance physics of many-body systems remains in
part an unsettled problem. This is the case for the class of models in
which $N$-component classical variables with $O(N)$ symmetric
interactions are coupled to a random field. Depending on whether the
coupling is linear or bilinear, the models belong to the 'random
field' (RF) or the 'random anisotropy' (RA) subclasses. Such models
with $N=1, 2$, or $3$ are relevant to describe a variety of systems
encountered in condensed matter physics or physical chemistry. To name
a few, one can mention dilute antiferromagnets in a uniform magnetic
field,\cite{belanger98} critical fluids and binary mixtures in
aerogels (both systems being modelled by the $N=1$ RF Ising
model),\cite{brochard83,degennes84, pitard95} vortex phases in
disordered type-II superconductors (described in terms of an elastic
glass model whose simplest version is the $N=2$ RF XY
model),\cite{giamarchi94,giamarchi95,giamarchi98,blatter94,nattermann00} amorphous magnets,
such as alloys of rare-earth compounds,\cite{harris73,dudka05} and
nematic liquid crystals in disordered porous media (described by $N=2$
or $N=3$ RA models).\cite{feldman01}

On the theoretical side the main questions raised about the
equilibrium behavior of such systems concern the nature and the
characteristics of the phases and of the phase transitions. It has
been shown by both heuristic and rigorous
methods\cite{imry75,bricmont87,imbrie84,larkin70,aizenman89} that the
lower critical dimension below which no long-range order is possible
is $\dlc=2$ for the RFIM and $\dlc=4$ for RF models with a continuous
symmetry ($O(N)$ with $N>1$). The same conclusion applies to RA models
with the restriction that only ferromagnetic (to use a magnetic
terminology) long-range order is forbidden below $\dlc=4$; another
type of long-range order associated to a spin-glass phase is still
possible.\cite{dudka05} Here, we only consider RA models with isotropic
distributions of the random anisotropies and with $N>1$; for
anisotropic distributions, long-range ferromagnetic ordering may still
occur below $\dlc=4$, whereas RA makes no real sense for $N=1$, the
model reducing then either to the random temperature Ising model or to
the pure Ising model depending on the details of the effective
hamiltonian.\cite{dudka05}

Two central issues remain under active debate. The first one is about
the so-called 'dimensional reduction' property. Standard perturbation theory
predicts to all orders that the critical behavior of an $O(N)$ model
in the presence of RF is the same as that of the pure model, \ie{} with
no RF, in two dimensions less.\cite{nattermann98} The same applies to
the RA$O(N)$M with $N>1$ near the paramagnetic-ferromagnetic
transition.\cite{fisher85} Dimensional reduction is known however  to
break down, its most striking failure being the prediction of a lower
critical dimension $\dlc=3$ for the RFIM in contradiction with the
exact result (see above). A proper description of the long-distance
behavior of RF and RA models must thus provide a way out of the
dimensional reduction.

The second issue concerns the phase diagram of the RF and RA models
with a continuous symmetry ($N>1$) in dimensions below $d=4$, which of
course are relevant to the physical situations. If long-range
ferromagnetism is forbidden, quasi-long-range order (QLRO), namely a
phase characterized by no magnetization and a power-law decrease of
the correlation functions at large distances, may still
exist.\cite{giamarchi94,giamarchi95,giamarchi98,blatter94,nattermann00,feldman00}
It has been shown that QLRO is absent for $N\geq3$ in the presence of
RF and for $N\geq10$ in the presence of RA;\cite{feldman00}
yet it has been argued that QLRO is present for $N=2$ in $d=3$, in
which case it corresponds to the 'Bragg glass' phase predicted for
vortices in disordered type-II
superconductors.\cite{giamarchi94,giamarchi95,giamarchi98}

We have recently proposed a coherent resolution of those issues based
on a non-perturbative (NP) functional renormalization group (FRG)
treatment.\cite{tarjus04,tissier06} This approach has allowed us to provide a
unified picture of ferromagnetism, QLRO and criticality in RF models
in the whole ($N$, $d$) diagram as well as a way to escape dimensional
reduction.

\begin{figure}[htbp]
  \centering
  \includegraphics[width=\linewidth]{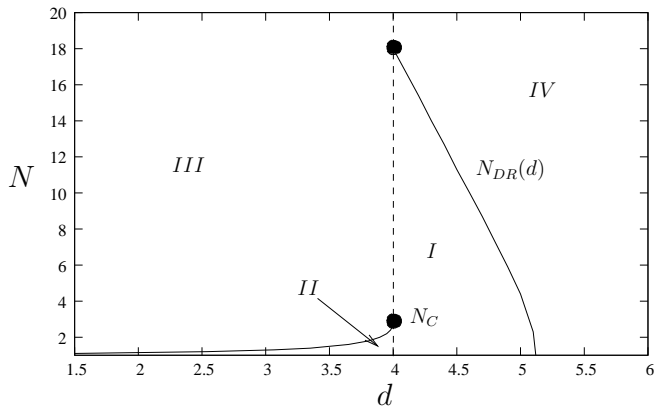}
  \caption{Nonperturbative FRG prediction for the $(N,d)$ phase
  diagram of the RF$O(N)$M. See text for comments.}
  \label{fig_phase_diag_np}
\end{figure}
The main findings\cite{tarjus04,tissier06} can be summarized on the
phase diagram of the RF$O(N)$M shown in
Figure~\ref{fig_phase_diag_np}. In region III, there are no phase
transitions and the system is always disordered (paramagnetic). In
regions I and IV, there is a second-order paramagnetic to
ferromagnetic transition and in region II, a second-order transition
between paramagnetic and QLRO phases. In all cases, the critical
behavior is controlled by a zero-temperature fixed point at which
temperature is formally irrelevant. At this fixed point the
\textit{renormalized} effective action is a nonanalytic function of
its arguments (the order parameter fields). Although present, the
nonanalyticity is weak enough in region IV to let the critical
exponents take their dimensional reduction value (corrections to
scaling may nonetheless differ from the dimensional reduction
predictions). In regions I and II the nonanalyticity takes the form of
a cusp in the renormalized second cumulant of the random field, which
leads to a complete breakdown of dimensional reduction. Finally, the
whole QLRO phase in region II is also controlled by a zero-temperature
fixed point characterized by a cusp.

There is undoubtedly room for improving the quantitative predictions
of our NP-FRG theory, in terms of both the number of observables
studied and, more importantly, of the accuracy of the (necessary)
approximations involved. (In addition, the NP-FRG study of the
RA$O(N)$M has not yet been completed.) The robustness of the proposed
scenario may however be tested by considering a \textit{perturbative}
FRG treatment of the models near $d=4$. Such a perturbative FRG has
been pioneered by D. Fisher\cite{fisher85,fisher86} and widely used to
study the statics and the depinning of elastic systems pinned by
quenched
disorder.\cite{giamarchi94,giamarchi95,giamarchi98,ledoussal02,ledoussal04}

At one-loop level, the flow equation for the renormalized second
cumulant of the disorder, first derived by D. Fisher\cite{fisher85}
for the RF and RA $O(N)$ models, has been studied by several
authors.\cite{feldman00,feldman02,tarjus04,tissier06,sakamoto06}
The results fit into the diagram displayed in
Figure~\ref{fig_phase_diag_np}, which should come as no surprise: the
flow equations obtained in our NP-FRG approach exactly reproduce the
1-loop result near $d=4$.\cite{tarjus04} Below, we give a survey of
the behavior of the RF$O(N)$M at 1-loop level in $d=4+\varepsilon$,
including some new results, as well as a study of the related
RA$O(N)$M.

To go beyond this first step, one must consider the next order in the
loop expansion. However, the technical difficulties are now much more
involved than at the 1-loop level. On top of the rapidly increasing
number of diagrams, diagrams which in the present case are
functionals, the nonanalytic character of the renormalized effective
action at $T=0$ leads to the appearance of 'anomalous' terms in the
diagramatics, whose evaluation is a priori ambiguous. A resolution of
the problem has been proposed for the simpler case of disordered
elastic systems by Le Doussal, Wiese and coworkers.\cite{ledoussal04}

A preliminary account of this work has been published in
Ref.~\onlinecite{tissier06}. An independent calculation has appeared
in Ref.~\onlinecite{ledoussal06}.

The rest of the paper is organized as follows. In section
\ref{sec_models} we present the RF and RA $O(N)$ models and their
nonlinear sigma versions appropriate to describe the long-distance
physics near the lower critical dimension of ferromagnetism, $d=4$. We
outline the perturbative FRG framework and the way to extract scaling
behavior and critical exponents. Section~\ref{sec_one-loop} is devoted
to an analysis of the 1-loop FRG equations at $T=0$ in $d=4+\epsilon$,
discussing the fixed points and their stability and contrasting the RF
and RA cases.  In section~\ref{sec_two_loops} we derive the FRG beta
functions at 2-loop order in $T=0$. We present the diagrammatic
representation and the method used to handle the apparent ambiguities
appearing in the formulation due to the nonanalytic character of the
renormalized dimensionless effective action. Proceeding in this way we
obtain a well-defined renormalized theory at 2-loop order. The
physical results obtained from solving the 2-loop FRG equations are
discussed in section~\ref{sec_fp_2-loop}, and we stress the new
features appearing at this order.  Several technical aspects of the work
are deferred to Appendices.

\section{Models and framework}
\label{sec_models}

\subsection{Models}

We consider the $O(N)$ model in the presence of RF or RA near
$d=4$. We stress again that $d=4$ is the lower critical dimension for
$N>1$ and for the paramagnetic to ferromagnetic transition. In a
manner similar to that developed for the pure model at low temperature
near $d=2$, the long-distance physics for weak disorder (which, we
recall, takes here the role played by low temperature in the pure
model, temperature being now irrelevant and eventually set to zero)
can be described in a field-theoretical setting by a nonlinear sigma
model with effective Hamiltonian
\begin{equation}
  \label{eq_ham_dis}
  \begin{split}
\mathcal H[\vect S]=\int d^d x\ &\frac 12(\nabla \vect S(\vect x))^2 -  
\sum_i(h^i(\vect x)+H \hat u^i) S^i(\vect x)\\&-\sum_{ij}\tau^{ij}(\vect x)S^i(\vect x)S^j(\vect x)
  \end{split}
\end{equation}
where the $N$-component spins $\vect{S}$ satisfy the fixed-length
constraint, $\mathbf{S}(\vect{x})^{2}=1$, and
$\vect{H}=H\vect{\hat{u}}$ is a uniform external magnetic field;
$\vect{h}(\vect{x})$ is a random magnetic field and
$\vect{\tau}(\vect{x})$ a second-rank random anisotropy tensor,
both with gaussian distributions characterized by zero means and
variances given by:
\begin{equation}
  \label{eq_cumulant_h}
  \overline{h^i(\vect x)h^j(\vect y)}=\Delta\ \delta_{ij}\ \delta(\vect x-\vect y)
\end{equation}
\begin{equation}
  \label{eq_cumulant_tau}
  \overline{\tau^{ij}(\vect x)\tau^{kl}(\vect y)}=\frac{\Delta_{2}}{2} \left( \delta_{ik}\delta_{jl}+\delta_{il}\delta_{jk}\right) \delta(\vect x-\vect y).
\end{equation}
Higher-order random anisotropies could be included as well. They will
indeed be generated in the perturbation expansion and the
renormalization group flow.\cite{fisher85} However, starting with only a
second-rank (or more generally an even-rank) random anisotropy, only
even-rank anisotropies will be generated. In what follows we will
therefore use the acronym RA to characterize models with even-rank
random anisotropies.

From the associated partition function,
\begin{equation}
  \label{eq_part_func}
  \mathcal Z=\int\mathcal D \vect S\ \delta(\vect S^2-1) \exp\left(-\frac 1 T \mathcal H[\vect S]\right ) ,
\end{equation}
one can obtain the free energy by averaging the logarithm of $\mathcal
Z$ over the quenched disorder. This is more conveniently performed by
introducing replicas $\vect {S}_{a}(\vect {x})$, $a=1,...,n$, which
leads after explicitly performing the average over the disorder to the
following 'replicated' effective Hamiltonian
\begin{equation}
  \begin{split}
    \mathcal H_n[\{\vect S_a\}]=\int d^dx&\sum_a \frac 12(\nabla \vect
    S(\vect x))^2 -
    \sum_aH \vect {\hat{u}}\cdot\vect S(\vect x)\\
    & -\frac 1{2T}\sum_{ab} R_0\left(\vect S_a(\vect x)\cdot\vect
      S_b(\vect x)\right)
  \end{split}
\end{equation}
with
\begin{equation}
  \label{eq_R_init}
  R_0(z)=\Delta z+ \Delta_2 z^2
\end{equation}
and $-1\leq z\leq+1$. The fluctuations around a fully ordered state in
which spins in all replicas align in the same direction are as usual
handled by splitting the replica field $\vect {S}_{a}(\vect {x})$
into a component collinear to the external field (and to the
magnetization),
$\Sigma_{a}(\vect {x})=\vect {S}_{a}(\vect {x}).\vect {\hat{u}}$,
and $N-1$ components orthogonal to it,
$\vect {\Pi}_{a}(\vect {x})=\vect {S}_{a}(\vect {x})-\Sigma_{a}(\vect {x})\vect {\hat{u}}$. By
using the relation between $\Sigma_{a}(\vect {x})$ and
$\vect {\Pi}_{a}(\vect {x})$ imposed by the unit-length constraint,
the replicated partition function can be expressed as a functional
integral over the ($N-1$)-component replica fields
$\vect {\Pi}_{a}(\vect {x})$,
\begin{equation}
  \label{eq_part_func_replica}
  \begin{split}
    \mathcal Z_n=\int&\prod_a \mathcal D \vect \Pi_a\
 \\\exp&\left(- \sum_a \mathcal S_1[\vect
      \Pi_a]+\frac 12 \sum_{ab} \mathcal S_2[\vect \Pi_a,\vect \Pi_b]+\cdots\right),
      \end{split}
\end{equation}
where the 1-replica and 2-replica parts of the action read
\begin{equation}
\label{eq_S1}
\begin{split}
  \mathcal S_1[\vect \Pi_a]=\frac 1T\int d^dx\Big\{ \frac 12(\nabla \vect
  \Pi_a)^2&+\frac {\left(\vect \Pi_a \cdot \nabla \vect
  \Pi_a)\right)^2}{2(1-\vect \Pi_a^2)} \\&-
  H \sqrt{1-\vect\Pi_a^2}\Big\}
\end{split}
\end{equation}
\begin{equation}
  \label{eq_S2}
  \begin{split}
    \mathcal S_2[\vect \Pi_a,\vect \Pi_b]=\frac 1{T^2}\int d^dx\ 
    R_0\Big(&\vect \Pi_a\cdot\vect
    \Pi_b+\\&\sqrt{1-\vect\Pi_a^2} \sqrt{1-\vect\Pi_b^2} \Big)
   \end{split}
\end{equation}
and the dots denote terms, such as those produced by the Jacobian of
the transformation from the $\vect {S}_{a}$'s to the
$\vect {\Pi}_{a}$'s and possible contributions involving more than
two replicas, that either do not contribute to the perturbation
expansion in the $T=0$ limit or turn out to be irrelevant within
conventional power counting.\cite{fisher85,ledoussal04}

From the logarithm of the partition function,
Eq.~(\ref{eq_part_func_replica}), one can obtain, by a Legendre
transform with respect to external sources coupled to the
($N-1$)-component replica fields $\vect {\Pi}_{a}(\vect {x})$, the
effective action $\Gamma_{n}\left[\left\lbrace \vect {\Pi}_{a}
  \right\rbrace \right]$ which is the generating functional of the
one-particle irreducible vertices for the $\vect {\Pi}_{a}$ fields and
from which all equilibrium observables can be derived. (The subscript
$n$ will be dropped in the following.)

\subsection{Perturbation theory and renormalization}
\label{sec_pert_renorm}

We proceed by calculating the effective action
$\Gamma\left[\left\lbrace \vect {\Pi}_{a} \right\rbrace \right]$
perturbatively in powers of the disorder correlator $R_{0}$, keeping
only terms that do not vanish in the limit $T=0$. The results so
obtained would however be singular, showing the standard ultra-violet
divergences as $\epsilon=d-4$ goes to zero. For instance, if we use
the dimensional regularization as a regularization scheme, the 1-loop
calculation brings in terms proportional to $1/\epsilon$. To cure this
problem, it is necessary to renormalize the theory by introducing in
the effective Hamiltonian 'counterterms' that are chosen to keep the
physical quantities finite.

Expressed in terms of dimensionless renormalized quantities at an
arbitrary momentum scale $\mu$, the 1- and 2-replica parts of the
action read:
\begin{equation}
\label{eq_S1_re}
\begin{split}
  \mathcal S_1[\vect \pi_a]=\frac {\mu^{d-2} Z_\Pi}{2Z_Tt}\int& d^dx\Big\{
 (\nabla \vect \pi_a)^2+Z_\Pi\frac {\left(\vect \pi_a \cdot \nabla \vect
      \pi_a)\right)^2}{1- Z_\Pi\vect \pi_a^2} \\&- 2Z_\Pi^{-1/2}Z_Th
  \sqrt{1-Z_\Pi\vect\pi_a^2}\Big\}
\end{split}
\end{equation}
\begin{equation}
  \label{eq_S2_re}
  \begin{split}
    \mathcal S_2[\vect \pi_a,\vect \pi_b]=\frac
    {\mu^{d}}{2Z_T^2t^2}\int &d^dx\ Z_{R}\Big(z_0=Z_\Pi\vect
    \pi_a\cdot\vect\pi_b+\\&\sqrt{1-Z_\Pi\vect\pi_a^2}
    \sqrt{1-Z_\Pi\vect\pi_b^2}\Big)
\end{split}
\end{equation}
where the dimensionless renormalized quantities are defined as
\begin{subequations}
\label{eq_renormalized}
\begin{align}
\label{eq_renorm_pi}\vect \Pi&=\sqrt {Z_\Pi} \vect \pi\\
\label{eq_renorm_T}  T&=\mu^{2-d}Z_Tt \\
\label{eq_renorm_H}  H&=Z_T Z_\Pi^{-\frac 12}h\\
\label{eq_renorm_R}R_0&=\mu^{4-d}Z_{R}
\end{align}
\end{subequations}
and $Z_{R}(z)$ is a functional of the renormalized dimensionless disorder
correlator $R(z)$, with its leading term equal to $R(z)$. The two
renormalization constants $Z_{T}$ and $Z_{\Pi}$ and the
renormalization function $Z_{R}(z)$ are chosen so that the
loopwise perturbative expansion of the effective action remains
finite. (We work in the minimal subtraction scheme in which the
counterterms contain only the singular parts necessary to make the
physical quantities finite.) In practice we compute the 2-point proper
vertex associated with the 1-replica part of the effective action,
$\Gamma_{1}^{(2)}(q)$, and the 2-replica part of the effective action,
$\Gamma_{2}$, both being evaluated for uniform configurations of the
$\vect \pi_{a}$ fields.

The perturbative expansion is organized about the free theory formed
by the quadratic part of the 1-replica action. The associated free
propagators are expressed in terms of the bare quantities as follows:
\begin{equation}
\label{eq_propag}
G_{ij}(q)= T \frac {\delta_{ij} -\Pi_i\Pi_j}{q^2+H/\Sigma}
\end{equation}

where $\Sigma=1-\vect \Pi^{2}$. Note that following Brezin and
Zinn-Justin\cite{brezin76} we keep an external magnetic field $H$
which allows to regularize the infrared divergences by giving a mass
to the Goldstone modes. Aside from this term, the action in Eqs.
(10,11) is $O(N)$ invariant. The loop expansion can be graphically
expressed in terms of 1-particle irreducible Feynman diagrams with
vertices coming from both the non-quadratic piece of the 1-replica
action and from the 2-replica action.

A difficulty of the present problem lies in the functional character
of the expansion, the 2-replica vertices involving the whole function
$R(z)$ and its derivatives. This is somewhat similar to the treatment
of disordered elastic systems,\cite{ledoussal04} with however the
additional complication that the 1-replica part is now nontrivial and gets
renormalized in a manner that couples to the renormalization of the
disorder. The details of the calculation as well as the method to
handle possibly anomalous terms appearing at 2-loop level when the
renormalized correlator of the disorder is nonanalytic will be
presented in Section~\ref{sec_two_loops}.

\subsection{FRG equations, critical exponents and correlation functions}

For the 1-replica, 2-point proper vertex and for the 2-replica
effective action (when both evaluated for uniform configurations of
the fields), the relation between the renormalized and the bare
theories is simply
\begin{align}
  \label{eq_gamma1bare}
  \Gamma_{1,\mu}^{(2)}(q;\vect \pi,t,h,R)&=Z_\Pi\ \Gamma_{1,\text B}^{(2)}(q;\vect \Pi,T,H,R_0)\\
  \label{eq_gamma2bare}
  \Gamma_{2,\mu}(z,t,h,R)&=\Gamma_{2,\text B}(z,T,H,R_0)
\end{align}
where $\text B$ denotes the bare theory.

The RG flow equations then result from the invariance of the bare
theory under a change of the momentum scale $\mu$, when $T, H$ and
$R_{0}$ are held fixed. Actually, we are only interested in the situation of zero temperature ($T=0$) and zero external field ($H=0$). We introduce
\begin{align}
  \label{eq_zeta_pi}
  \zeta_\Pi=\mu\partial_\mu \log Z_\Pi|_{R_0}\\
\label{eq_zeta_T}
  \zeta_T=\mu\partial_\mu \log Z_T|_{R_0}
\end{align}
and 
\begin{equation}
  \label{eq_betaR}
  \beta_R(z)=-\mu\partial_\mu R(z)|_{R_0},
\end{equation}
where we have implicitly set $T=H=0$. As an illustration, the flow of
the 1-replica proper vertex, $t \Gamma_{1,\mu}^{(2)}(q)$, when
$H=h=0$, $T=t=0$, $\vect \Pi=\vect \pi=\vect {0}$ is derived as
\begin{equation}
  \label{eq_flot_R}
  \begin{split}
    \Big[\mu\partial_\mu +(2&-d+\zeta_T-\zeta_\Pi)-\\&\int_{-1}^1dz'\ 
    \beta_R(z')\frac{\delta\ }{\delta
      R(z')}\Big](t \Gamma_{1,\mu}^{(2)}(q))=0
      \end{split}
\end{equation}
where the long-distance physics is now obtained when $\mu\rightarrow 0$.

The scaling behavior and the critical exponents of the physical
quantities can be obtained from the fixed-point solutions and the
properties of the flow near the fixed points. In particular, the
exponents $\eta$ and $\bar{\eta}$ that characterize the power-law
decay of the 2-point correlation functions at the critical point for
small $q$,
\begin{equation}
  \label{eq_correl_conn}
  \overline{\langle\vect S(-\vect q)\cdot \vect S(\vect q)\rangle}- \overline{
    \langle\vect S(-\vect q)\rangle\cdot \langle\vect
    S(\vect q)\rangle}\sim q^{-(2-\eta)}
\end{equation}
\begin{equation}
  \label{eq_correl_disc}
  \overline{
    \langle\vect S(-\vect q)\rangle\cdot \langle\vect
    S(\vect q)\rangle}-\overline{
    \langle\vect S(-\vect q)\rangle}\cdot\overline{
    \langle\vect S(\vect q)\rangle} \sim q^{-(4-\bar \eta)},
\end{equation}
and the exponent $\theta$ associated with the temperature, $t=
\mu^{\theta} T$, are given by
\begin{align}
  \label{eq_eta}
  \eta&=\zeta_{\Pi\ast}-\zeta_{T\ast}\\
\label{eq_etab}
  \bar\eta&=4-d+\zeta_{\Pi\ast}\\
\label{eq_theta}
\theta&=d-2-\zeta_{T\ast}=2-\bar\eta+\eta
\end{align}
where $\zeta_{\Pi\ast}$ and $\zeta_{T\ast}$ are the fixed-point values
of Eqs. (\ref{eq_zeta_pi},\ref{eq_zeta_T}). Provided $\theta>0$, the fixed point indeed occurs at
zero renormalized temperature.

Before closing this section, it is worth recalling an inequality for
the correlation functions in the present models, which turns into an
inequality between critical exponents. In the RF case, the result is
due to Schwartz and Soffer,\cite{schwartz85} who have proven that the
$\vect q$ Fourier component of the 'connected' pair correlation
function, $\overline{\left\langle \vect {S}(-\vect {q}) \vect {.}
    \vect {S}(\vect {q})\right\rangle} - \overline{\left\langle \vect
    {S}(-\vect {q})\right\rangle \vect {.} \left\langle \vect
    {S}(\vect {q})\right\rangle}$, is always less than the square root
of the $\vect {q}$ component of the 'disconnected' pair correlation
function, $\overline{\left\langle \vect {S}(-\vect {q})\right\rangle
  \vect {.} \left\langle \vect {S}(\vect {q})\right\rangle} -
\overline{\left\langle \vect {S}(-\vect {q})\right\rangle} \vect {.}
\overline{\left\langle \vect {S}(\vect {q})\right\rangle}$, up to an
irrelevant multiplicative constant. As a consequence, one must have
$\bar{\eta}\leq 2\eta$.

The RA case is different and has been considered by
Feldman.\cite{feldman00} In this model indeed, the
randomness couples to a composite field that is bilinear in the spin
variables (see Eq.~(\ref{eq_ham_dis})). As a consequence, the
inequality now applies to the connected and disconnected correlation
functions of the composite (bilinear) field.  One can define new
critical exponents, $\eta_{2}$ and $\bar{\eta}_{2}$, for those
correlation functions,
\begin{equation}
  \label{eq_correl_conn_ra}
  \overline{\langle\vect m(-\vect q)\cdot \vect m(\vect q)\rangle}- \overline{
    \langle\vect m(-\vect q)\rangle\cdot \langle\vect
    m(\vect q)\rangle}\sim q^{-(2-\eta_{2})}
\end{equation}
\begin{equation}
  \label{eq_correl_disc_ra}
  \overline{
    \langle\vect m(-\vect q)\rangle\cdot \langle\vect
    m(\vect q)\rangle}-  \overline{
    \langle\vect m(-\vect q)\rangle}\cdot \overline{
    \langle\vect m(\vect q)\rangle} \sim q^{-(4-\bar{\eta}_{2})}
\end{equation}
where $m^i(\vect {x})=S^{i}(\vect {x})^{2}-(1/N)$. The inequality
between the correlation functions then imposes that
$\bar{\eta}_{2}\leq 2\eta_{2}$. However, the exponents $\eta$ and
$\bar{\eta}$ are no longer constrained by the usual Schwartz-Soffer
inequality. The exponents $\eta_{2}$ and $\bar{\eta}_{2}$ are also
expressable in terms of fixed-point quantities.

In the following we first analyze the FRG equations obtained at 1-loop
order near $d=4$.

\section{Analysis of the 1-loop FRG equations in $d=4+\epsilon$}
\label{sec_one-loop}

\subsection{1-loop beta function in $d=4+\epsilon$}
\label{sec_one-loop_a}

The beta function for the renormalized correlator of the disorder
$R(z)$ at zero temperature has been obtained by Fisher at the 1-loop
level in $d=4+\epsilon$.\cite{fisher85} it reads
\begin{equation}
  \label{eq_beta_1_loop}
  \begin{split}
  \beta_R(z)&=-\mu\partial_\mu R(z)=-\epsilon
  R(z)+C\Big(2(N-2)R(z)R'(1)\\&+\frac 12
  (N-2+z^2) R'(z)^2-z(1-z^2)R'(z)R''(z)\\ &+\frac 12
  (1-z^2)^2 R''(z)^2-(N-1)z R'(1)R'(z)\\&+(1-z^2)R'(1)R''(z)\Big)
      \end{split}
\end{equation}
where $C=1/(8\pi^{2})$. The above expression is valid for both the RF
and the RA models. The only difference is the additional inversion
symmetry present in the latter: $z$ goes from $-1$ to $+1$ in all
cases, but in the RA model, $R(-z)=R(z)$.

At the same 1-loop level, the critical exponents $\eta, \bar{\eta}$
and $\eta_{2}, \bar{\eta}_{2}$ defined in
Eqs.~(\ref{eq_correl_conn},\ref{eq_correl_disc}) and
(\ref{eq_correl_conn_ra},\ref{eq_correl_disc_ra}) are given by
\cite{fisher85,feldman00}
\begin{align}
  \label{eq_eta_1l}
  \eta&=C R_{\ast}'(1)\\
\label{eq_etab_1l}
  \bar\eta&=-\epsilon+(N-1)C R_{\ast}'(1)\\
  \label{eq_eta2_1l}
  \eta_2&=(N+2)C R_{\ast}'(1)\\
\label{eq_etab2_1l}
  \bar\eta_2&=-\epsilon+2NC R_{\ast}'(1)
\end{align}
where the star indicates a fixed-point solution.

For studying the fixed points and their stability it is convenient to
introduce $\widetilde{R}(z)=(C/\epsilon)R(z)$ (the renormalized disorder
is of order $\epsilon$ at the putative fixed points) and to consider
the beta function for its derivative,
\begin{equation}
  \label{eq_beta_Rp}
  \begin{split}
    \epsilon^{-1}&\beta_{\widetilde R'}(z)=-\widetilde R'(z)+z
    \widetilde R'(z)^2+ (N-3) \widetilde R'(z)\widetilde
    R'(1)\\&+\left(N-3+4 z^2\right) \widetilde R'(z)\widetilde R''(z)-
    (N+1) z \widetilde R'(1)\widetilde R''(z)\\&-z \left(1-z^2\right)
    \widetilde R'(z)\widetilde R'''(z)+\left(1-z^2\right) \widetilde
    R'(1)\widetilde R'''(z)\\&-3 z \left(1-z^2\right)\widetilde
    R''(z)^2 +\left(1-z^2\right)^2 \widetilde R''(z) \widetilde
    R'''(z)
   \end{split}
\end{equation}

It is illustrative to write down the 1-loop beta functions for the
first derivatives $\widetilde{R}'(z=1)$ and $\widetilde{R}''(z=1)$, assuming
that $\widetilde{R}(z)$ is at least twice continuously differentiable
around $z=1$. ($z=1$ corresponds to the situation where the spins in
the two considered replicas become equal,
$\vect {S}_{a}=\vect {S}_{b}$.) The expressions are
\begin{equation}
  \label{eq_beta_Rp1}
  \epsilon^{-1}\beta_{\widetilde R'(1)}=-\widetilde
  R'(1)+(N-2)\widetilde R'(1)^2 
\end{equation}
\begin{equation}
  \label{eq_beta_Rs1}
  \begin{split}
  \epsilon^{-1}\beta_{\widetilde R''(1)}=-(-1+&6 \widetilde
  R'(1))\widetilde  R''(1)\\&+(N+7)\widetilde R''(1)^2+\widetilde  R'(1)^2 .
      \end{split}
\end{equation}

If $\widetilde{R}(z)$ is analytic around $z=1$, the beta functions for the
higher derivatives evaluated at $z=1$ can be derived as well. As noted
by Fisher,\cite{fisher85} the expression for the $p$th derivative only
involves derivatives of lower or equal order (and for $p\geq3$ the
beta function is linear in the $p$th derivative). This structure
allows an iterative solution of the fixed-point equation, provided of
course that $\widetilde{R}(z)$ has the required analytic property.

The fixed points corresponding to the zeros of Eq.~(\ref{eq_beta_Rp1})
are $\widetilde{R}_{\ast}'(1)=0$ (stable) and
$\widetilde{R}_{\ast}'(1)=1/(N-2)$ (unstable with an eigenvalue
$\Lambda_{1}=\epsilon$). The latter fixed point leads to the
dimensional-reduction value of the critical exponents, \eg,
$\eta=\bar{\eta}=\epsilon/(N-2)$, $\nu=1/\Lambda_{1}=1/\epsilon$. The
second expression, Eq.~(\ref{eq_beta_Rs1}), has then (two) nontrivial
zeros only if $N\geq18$: the fixed point with
$\widetilde{R}_{\ast}''(1)=\frac{(N-8)+\sqrt{(N-2)(N-18)}}{2(N-2)(N+7)}$
is unstable ($\Lambda_{2}=\sqrt{\frac{(N-18)}{(N-2)}}\epsilon$)
whereas that with
$\widetilde{R}_{\ast}''(1)=\frac{(N-8)-\sqrt{(N-2)(N-18)}}{2(N-2)(N+7)}$
has a negative second eigenvalue,
$\Lambda_{2}=-\sqrt{\frac{(N-18)}{(N-2)}}\epsilon$.

This little exercise already shows that no nontrivial fixed-point
function $\widetilde{R}_{\ast}(z)$, twice differentiable in $z=1$, can
exist for $N<18$. Actually, one finds that there is a finite range of
initial conditions for $\widetilde{R}'(1)$ for which, no matter what
one chooses for its initial value, the RG flow for
$\widetilde{R}''(1)$ leads to a divergence at a \textit{finite} scale
$\mu$. The solution to this problem has been known for some
time:\cite{fisher86} the proper fixed point controlling the critical
behavior must be nonanalytic around $z=1$, with
$\widetilde{R}_{\ast}'(z)$ having a cusp, \ie{}, a term proportional
to $\sqrt{1-z}$ when $z\rightarrow1$, at least when $N<18$.

We now consider, separately and in more detail, the results for the
RF$O(N)$M and the RA$O(N)$M.

\subsection{RF $O(N)$ model}

Numerical solutions of the fixed-point equation,
$\beta_{\widetilde{R}'}(z)=0$, have been given by
Feldman\cite{feldman02} for $N=3, 4, 5$ and by us for general values
of $N$.\cite{tarjus04} Some analytical results can also be derived
and will be discussed at the end of this subsection. The picture one
gets from the numerical solutions is that the long-distance physics of
the RF $O(N)$ model near $d=4$ drastically depends on whether $N$ is
above or below two distinct critical values: $N_{DR}=18$ and
$N_{c}=2.8347...$.

\begin{figure}[b]
  \centering
  \includegraphics[width= \linewidth]{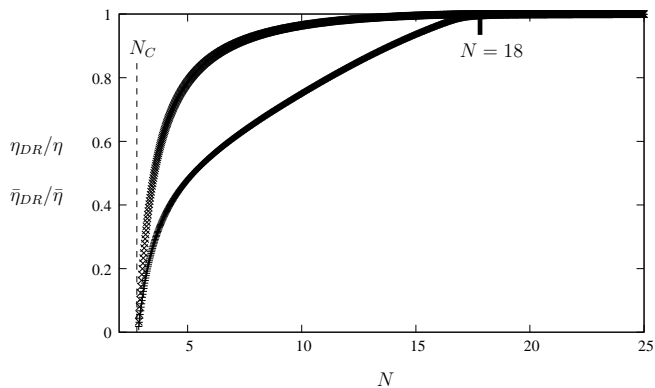}
  \caption{Ratios $\eta_{DR}/\eta$ (upper curve) and
    $\etab_{DR}/\etab$ (lower curve) vs $N$ for the RFO(N$>$2)M at
    first order in $\epsilon=d-4$ (the dimensional-reduction value for
    the exponents is $\eta_{DR}=\etab_{DR}=\epsilon/(N-2)$). The
    critical value of $N$ at which both $\eta$ and $\etab$ diverge is
    $N_c =2.8347...$.}
  \label{fig_eta_etab_rfon_ng}
\end{figure}

\begin{figure}[b]
  \centering
  \includegraphics[width= \linewidth]{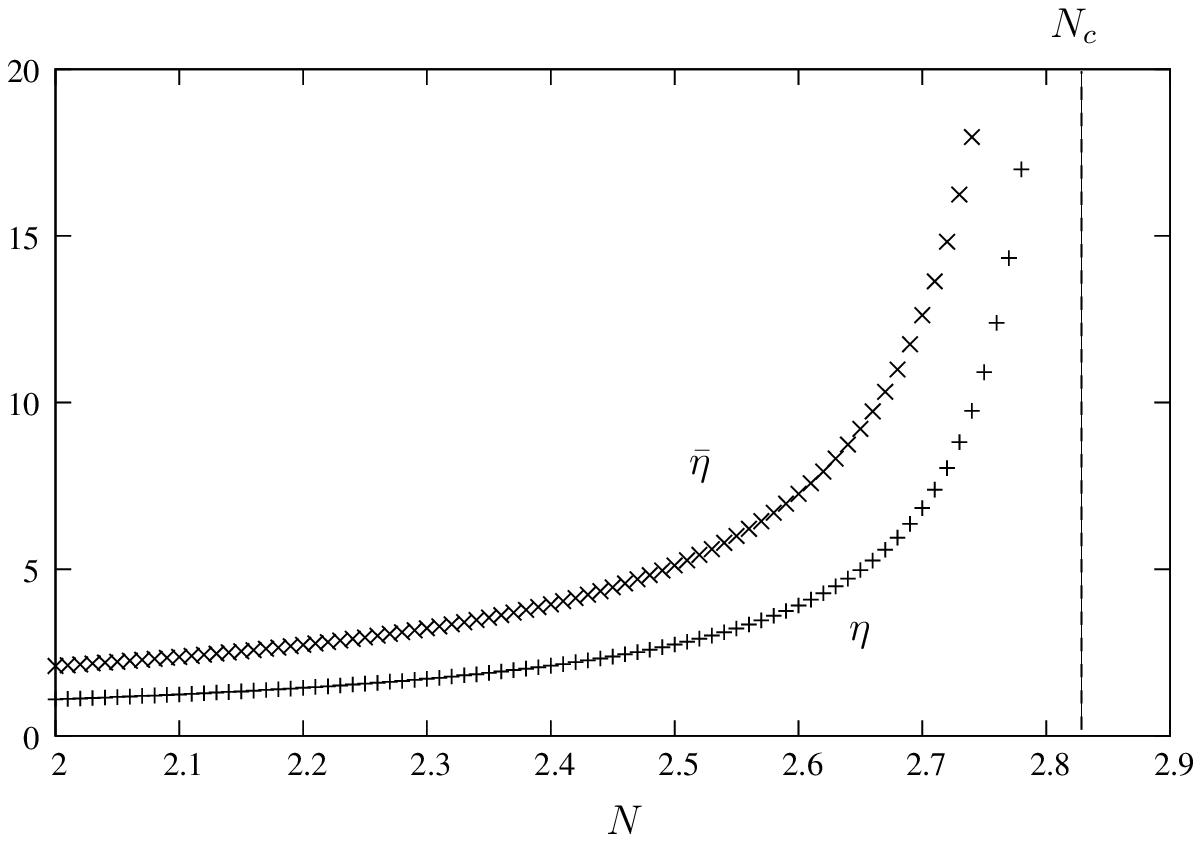}
  \caption{Exponents $\eta$ and $\etab$ (divided by $\epsilon$)
    characterizing the power-law decay of the pair correlations in the
    QLRO phase of the RF$O(N)$M for $N<N_c =2.8437...$ (first order in
    $\epsilon=d-4$).}
  \label{fig_eta_etab_rfon_np}
\end{figure}

The value $N_{DR}=18$ separates a region in which
$\widetilde{R}_{\ast}'(z)$ at the critical, \ie{}, once unstable, fixed
point has a cusp ($N<N_{DR}$) from a region ($N>N_{DR}$) where
$\widetilde{R}_{\ast}'(z)$ has only a weaker nonanalyticity, a 'subcusp'
in $(1-z)^{\alpha(N)}$ with $\alpha(N)$ a noninteger strictly larger
than $1$.\footnote{For some values of $N$, $\alpha(N)$ may
  accidentally be an integer, but logarithmic corrections should be
  present in this case.} As already mentioned, the occurence of a cusp
changes the values of $\eta$ and $\bar{\eta}$ from their
dimensional-reduction value,
$\eta_{DR}=\bar{\eta}_{DR}=\epsilon/(N-2)$. On the other hand, the
weaker nonanalyticity occuring for $N>18$ does not alter the
fixed-point value of $\widetilde{R}_{\ast}'(1)$ from that obtained from
Eq.~(\ref{eq_beta_Rp1}); this leads to $\eta=\bar{\eta}=\eta_{DR}$.
This is illustrated in Figure~\ref{fig_eta_etab_rfon_ng} where we plot
$\eta_{DR}/\eta$ and $\bar{\eta}_{DR}/\bar{\eta}$ as a function of
$N$.
On the other hand, at $N_{c}=2.834...$, the perturbative 'cuspy'
fixed point describing the paramagnetic to ferromagnetic critical
point when $\epsilon>0$ disappears ($\eta$ and $\bar{\eta}$ diverge as
$N\rightarrow N_{c}^{+}$, see Figure~\ref{fig_eta_etab_rfon_ng}). Below
$N_{c}$ an attractive cuspy fixed point appears for $\epsilon<0$ that
now describes a whole phase with QLRO. The exponents $\eta$ and
$\bar{\eta}$ characterizing this QLRO phase are plotted versus $N$ in
Figure~\ref{fig_eta_etab_rfon_np}.

At the critical value $N_{c}$, the beta function for $R'(z)$ (unscaled
by $\epsilon$) in exactly $d=4$ has a (cuspy) fixed-point solution
$R_{\ast}'(z)$ for any arbitrary value of the renormalized disorder
strength $R_{\ast}'(1)$. We have noted in
Ref.~\onlinecite{tissier06} that the situation bears some similarity
with the pure $O(N)$ model near $d=2$. There, the critical value
$N_{c}$ below which a QLRO phase may occur for $\epsilon<0$ is
$N_{c}=2$, and for $N_{c}=2$ and $d=2$ the beta function for the
temperature identically vanishes, independently of the value of the
temperature.  The singular point ($N_{c}=2$, $d=2$) is characterized
by the existence of a Kosterlitz-Thouless transition. One may then
wonder whether the singular point of the RF$O(N)$M ($N_{c}=2.8347...$,
$d=4$), despite the absence of the Abelian property specific to the
$O(N=2)$ model, also possesses a Kosterlitz-Thouless transition. This
point will be adressed below with the help of the 2-loop calculation.

We now complement this numerical study by providing some analytical
results. We first show that for $d>4$, the critical point is always
characterized by a correlation-length exponent $\nu$ which is equal
(at one loop) to its dimensional-reduction value,
$\nu_{DR}=1/\epsilon$. The eigenvalue equation obtained by linearizing
the beta function, Eq.~(\ref{eq_beta_Rp}), for a small deviation from
the fixed-point solution,
$\delta(z)=\widetilde{R}'(z) - \widetilde{R}_{\ast}'(z)$, is
\begin{equation}
  \label{eq_stab}
  \begin{split}
    &\frac \Lambda \epsilon\delta(z)=\delta (z) \Big(\widetilde
    R_{\ast}'''(z) z^3+2 \widetilde R_{\ast}'(z) z-\widetilde R_{\ast}'''(z) z\\&+N
    \widetilde R_{\ast}'(1)-3 \widetilde R_{\ast}'(1)+\left(4 z^2+N-3\right)
    \widetilde R_{\ast}''(z)-1\Big) \\&+\delta (1) \Big((N-3) \widetilde
    R_{\ast}'(z)-(N+1) z \widetilde R_{\ast}''(z)-\\&\left(z^2-1\right) \widetilde
    R_{\ast}'''(z)\Big) +\delta '(z) \Big(\widetilde R_{\ast}'''(z) z^4-2
    \widetilde R_{\ast}'''(z) z^2\\&-N \widetilde R_{\ast}'(1) z-\widetilde R_{\ast}'(1)
    z+6 \left(z^2-1\right) \widetilde R_{\ast}''(z) z+\\&\left(4 z^2+N-3\Big)
      \widetilde R_{\ast}'(z)+\widetilde R_{\ast}'''(z)\right) -
    \left(1-z^2\right)\\& \left(z \widetilde R_{\ast}'(z)-\widetilde
      R_{\ast}'(1)-\left(1-z^2\right) \widetilde R_{\ast}''(z)\right) \delta ''(z) .
  \end{split}
\end{equation}
By substituting $\delta(z)=\widetilde{R}_{\ast}'(z)$ in the above
equation, one can easily check that the fixed-point solution is also
solution of the eigenvalue equation with a positive eigenvalue
$\Lambda_{1}=\epsilon$ (from which, $\nu=\nu_{DR}=1/\epsilon$). This
result is independent of the analytic or nonanalytic character of
$\widetilde{R}_{\ast}'(z)$. For $d<4$, $\widetilde{R}_{\ast}'(z)$ is also
solution of Eq.~(\ref{eq_stab}) with $\Lambda_{1}=\epsilon$, but the eigenvalue
is now negative, which allows the fixed point to be fully attractive.

For $N>18$ it is possible to adapt Fisher's arguments concerning the
hierarchy of flow equations for the successive derivatives of
$\widetilde{R}_{\ast}(z)$ evaluated at $z=1$.\cite{fisher85} (In his article
however, Fisher did not envisage nonanalytic fixed-point solutions.)
As explained above, a fixed point with a well defined second
derivative and an associated negative eigenvalue ($\Lambda_{2}<0$) can
be found for $N>18$. (Note that cuspy fixed points are also present,
but they are more than once unstable and correspond to putative
multicritical behavior; a detailed analysis of the fixed points and
their stability in the $N\rightarrow \infty$ limit has been recently
provided by Sakamoto et al.\cite{sakamoto06}) The flow equations for the
higher derivatives are linear, namely,
\begin{equation}
  \label{eq_flow_rp}
  \begin{split}
-\mu\partial_\mu\widetilde R^{(p)}(1)=&-\Lambda_p(\widetilde R'(1),
  \widetilde R''(1))\  \widetilde R^{(p)}(1)\\&+\mathcal F_p(\widetilde
  R'(1), \widetilde R''(1),\cdots, \widetilde R^{(p-1)}(1)),
    \end{split}
\end{equation}
provided of course that the $p$th derivative is well defined in
$z=1$. If $\widetilde{R}'(1)$ and $\widetilde{R}''(1)$ are chosen equal to
their fixed-point values, $\widetilde{R}_{\ast}'(1)=1/(N-2)$ and
$\widetilde{R}_{\ast}''(1)=\frac{(N-8)-\sqrt{(N-2)(N-18)}}{2(N-2)(N+7)}$,
one finds
\begin{equation}
  \label{eq_eigenvalue_p}
  \begin{split}
\Lambda_{p\ast}&=\frac\epsilon{N-2}\Big[2p^2-(N-1)p+(N-2)+\\&\frac{p(N-5+6p)}
  {2(N+7)}(N-8-\sqrt{(N-2)(N-18)})\Big] .
   \end{split}
\end{equation}

For a given $N$, $\Lambda_{p\ast}$ monotonically increases with $p$,
so that there exists an integer value $p_{\sharp}(N)$ such that
$\Lambda_{p_{\sharp}(N)\ast}<0$ and $\Lambda_{p_{\sharp}(N)+1\ast}>0$.
Starting with an analytic bare action, a fixed point value is reached
(provided $\widetilde{R}'(1)$ is appropriately tuned) at which the
first $p_{\sharp}(N)$ derivatives of $\widetilde{R}_{\ast}(z)$ are
well defined in $z=1$. The RG flow for the ($p_{\sharp}(N)+1$)th
derivative on the other hand goes to infinity, but only in the limit
$\mu \rightarrow 0$. (This is to be contrasted with the situation for
$N<18$ in which the second derivative $\widetilde{R}''(1)$ diverges at
a finite scale $\mu$, due to the nonlinear nature of the corresponding
beta function.) As a consequence, the ($p_{\sharp}(N)+1$)th derivative
of $\widetilde{R}_{\ast}(z)$ is not defined in $z=1$, and there must
be a nonanalyticity in $\widetilde{R}_{\ast}'(z)$ of the form
$(1-z)^{\alpha(N)}$ with $p_{\sharp}(N)-1<\alpha(N)<p_{\sharp}(N)$.

\begin{figure}[ht]
  \centering
  \includegraphics[width= \linewidth]{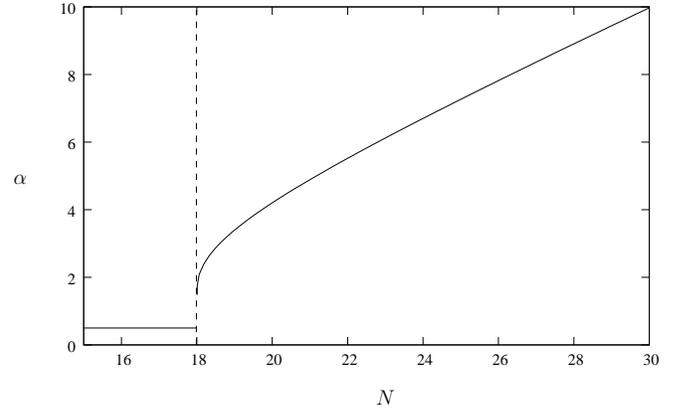}
  \caption{Exponent $\alpha(N)$ characterizing the nonanalyticity
    $(1-z)^{\alpha(N)}$ in $R'(z)$ for the RF$O(N)$M in $d=4+\epsilon$
    (at 1 loop). For $N<18$ $\alpha=1/2$ whereas
    $\alpha(N\to18^+=3/2$.}
  \label{fig_nonanalyticity}
\end{figure}

It is easy to show that the beta function for the coefficient, say
$a$, of the $(1-z)^{\alpha(N)}$ term is equal to
$\beta_a=\Lambda_{\alpha(N)+1\ast}a$, where
$\Lambda_{\alpha(N)+1\ast}$ is given by Eq. (\ref{eq_eigenvalue_p})
with $p$ replaced by the noninteger $\alpha(N)+1$. The beta function
is equal to zero with a nontrivial $a\neq 0$ if and only if
$\Lambda_{\alpha(N)+1}\ast=0$. This selects the form of the
nonanalyticity of the fixed-point solution around $z=1$. As noticed in
our previous work,\cite{tarjus04} the non-analyticity goes as
$N/2+O(1)$ at large $N$ (a behavior that cannot be captured in a $1/N$
expansion, see \eg{} Ref.~\onlinecite{sakamoto06}). However, contrary to what
stated in Ref.~\onlinecite{tarjus04}, the exponent of the subcusp increases
continuously with N when $N>18$, as illustrated in
Figure~\ref{fig_nonanalyticity}.

Finally, we close this survey of the RF$O(N)$M at 1-loop order by
pointing out that for $N=2$, the FRG equations, Eqs.
(\ref{eq_beta_1_loop}-\ref{eq_etab_1l}), exactly reduce to those of a
periodic elastic system, with a one-component displacement field,
pinned by disorder.  This is more easily seen by switching variables
from $z$ to an angle $\phi=\cos^{-1}(z)$. A cuspy fixed-point solution
(with $\eta=|\epsilon|\pi^2/9$ and $\bar{\eta}=|\epsilon|(1+\pi^2/9)$)
describing a QLRO phase (a 'Bragg glass') when $d<4$ is then
analytically obtained (compare with Refs.
\onlinecite{giamarchi94},\onlinecite{giamarchi95},\onlinecite{giamarchi98}).

\subsection{RA $O(N)$ model}
\label{sec_ra}

Overall, the picture of the long-distance physics of the RA $O(N)$
model that one gets from solving the 1-loop FRG equations is very
similar to that of the RF $O(N)$ model. This conclusion, we recall,
comes with the proviso that one focuses on weak disorder (working near
the lower critical dimension of the paramagnetic to ferromagnetic
transition) and that the possible spin-glass ordering which may occur
at stronger (finite) disorder is not considered. The main difference
with the behavior of the RF $O(N)$ model then lies in the critical
values $N_{DR}$ and $N_{c}$: $N_{c}$ is found to be $9.4412...$; on
the other hand, $N_{DR}=\infty$, which means that, contrary to the
RF case, a cusp appears for all values of $N$ and dimensional
reduction always breaks down completely.

To show that the fixed-point solution $\widetilde{R}_{\ast}'(z)$ has
always a cusp, it is instructive to go back to the beta functions for
the first two derivatives $\widetilde{R}'(1)$ and
$\widetilde{R}''(1)$, Eqs.~(\ref{eq_beta_Rp1},\ref{eq_beta_Rs1}),
assuming that there is no cusp, $(1-z)^{\alpha}$ with $0<\alpha<1$, in
$\widetilde{R}'(z)$. The RA model having the additional inversion
symmetry, $\widetilde{R}(-z)=\widetilde{R}(z)$, it is convenient to
rewrite $\widetilde{R}(z)=(1/2)\widetilde{S}(z^{2})$. From Eqs.
(\ref{eq_beta_Rp1},\ref{eq_beta_Rs1}) one obtains the flow equations for $\widetilde{S}'(1)$ and
$\widetilde{S}''(1)$. A nontrivial fixed point of the beta function
for $\widetilde{S}'(1)=\widetilde{R}'(1)$ is again
$\widetilde{R}_{\ast}'(1)=1/(N-2)$ and the associated eigenvalue is
positive ($\Lambda_{1}=\epsilon$), so that the fixed point can only be
reached if one tunes the initial value to be exactly $1/(N-2)$. When
doing so, the flow equation for $\widetilde{S}''(1)$ can now be
written as
\begin{equation}
\label{eq_flowSs}
\begin{split}
  -\epsilon^{-1}\mu\partial_\mu\widetilde S&''(1)=\frac 1{2(N-2)^2}+\\&2(N+7)(\widetilde
  S''(1) -\widetilde S''_+)(\widetilde
  S''(1) -\widetilde S''_-)
  \end{split}
\end{equation}
where $\widetilde{S}_{\pm}''=-\left[
\frac{(N+22)\mp\sqrt{(N-2)(N-18)}}{2(N-2)(N+7)}\right] $ are both
strictly negative for any finite value of $N$. If one starts with a
value of $\widetilde{S}''(1)$ that is positive or even zero, which is
indeed a physical requirement since the bare disorder correlator is of
the form $\widetilde{\Delta}_{2}z^{2}$ plus possible higher-order even
powers of $z$ associated with even-rank anisotropies, the beta
function, \ie{}, the right-hand side of Eq.~(\ref{eq_flowSs}), stays
strictly positive. As a result, $\widetilde{S}''(1)$ diverges, and it
actually diverges at a finite scale $\mu$. This is of course
incompatible with the hypothesis that the fixed point has a well
defined second derivative $\widetilde{S}_{\ast}''(1)$: a cusp must appear
along the flow. Note that this reasoning is completely independent of
$N$.

The limit $N\rightarrow \infty$ is however somewhat special. Looking
for $\widetilde{S}(z^{2})=O(1/N)$ and taking the $N\rightarrow \infty$
limit in Eq.~(\ref{eq_flowSs}), one finds
\begin{equation}
\label{eq_flowSs_inf}
\begin{split}
  -\epsilon^{-1}\mu\partial_\mu(N \widetilde S''(1))_\infty=2(N &\widetilde
  S''(1))_\infty\\&((N \widetilde
  S''(1))_\infty+1) .
  \end{split}
\end{equation}

A fixed point with $\widetilde{S}_{\ast}''(1)=0$, although having a second
positive eigenvalue, can still be reached from an inital condition
with $\widetilde{S}''(1)=0$: this analytic fixed point is the one found by
a direct analysis of the $N\rightarrow \infty$ saddle-point equation
of an RA model with \textit{only} a second-rank anisotropy,
$\Delta_{2}z^{2}$. It is, however, unstable to the introduction of
higher-order anisotropies (and of course never stable when $N$ is
large but finite).

One can actually find the cuspy fixed-point solution in the large $N$
limit. It is convenient to change variable from $z$ to
$\phi=\cos^{-1}(z)$ and define
$\widetilde{T}(\phi)=(N-2)\widetilde{R}(z)$. $\widetilde{T}(\phi)$ must be an
even function of $\phi$ and the symmetry $\widetilde{R}(-z)=\widetilde{R}(z)$
translates into $\widetilde{T}(\pi-\phi)=\widetilde{T}(\phi)$, so that it is
sufficient to consider $\phi$ in the interval $[0,\frac{\pi}{2}]$. The
beta function for $\widetilde{T}'(\phi)$ in the large $N$ limit is given
by
\begin{equation}
  \label{eq_sp_inf}
  \begin{split}
    \epsilon^{-1}&\beta_{\widetilde T'}(\phi)=-\widetilde T'(\phi)-\frac {\cos
      (\phi )}{\sin^3(\phi)} \widetilde T'(\phi )^2 + \\&\frac{\widetilde T'(\phi )}{\sin
      ^{2}(\phi )} \left(\cos (2 \phi ) \widetilde T''(0)+\widetilde T''(\phi )\right)
    -\\&\frac{\cos (\phi)}{\sin (\phi)} \widetilde T''(0) \widetilde T''(\phi
    )+\frac{ \widetilde T'''(\phi )}{N-2}\left(\widetilde T''(\phi)-\widetilde T''(0 )\right)
  \end{split}
\end{equation}
where the last term can be dropped in the large $N$ limit. Details on
the derivation of the solution and on the stability analysis are
provided in Appendix A. Here, we only quote the result:
\begin{equation}
  \label{eq_Sp_fp}
  \widetilde
  T_{\ast}'(\phi)=-3\sin\phi \cos\left(\frac{\pi+|\phi|}{3}\right) +
  \mathcal O\left(\frac 1N\right) .
\end{equation}

The $1/N$ correction can also be analytically obtained and is given in
Appendix~\ref{sec_ra_large_n}. One can see that $\widetilde{T}_{\ast}'(\phi)$ in Eq.~(\ref{eq_Sp_fp}) satisfies the
symmetry requirement around $\frac{\pi}{2}$, since
$\widetilde{T}_{\ast}'(\frac{\pi}{2})=0$, and has a cusp in
$\vert\phi\vert$ (\ie{}, in $\sqrt{1-z}$) as $\phi$ goes to zero (and
$z$ goes to 1). The fixed point is once unstable with, as shown in the
previous subsection, $\Lambda_{1}=\epsilon$; hence
$\nu=\nu_{DR}=1/\epsilon$. The critical exponents $\eta, \bar{\eta}$
and $\eta_{2}, \bar{\eta}_{2}$ are obtained from
$\widetilde{R}_{\ast}'(1)=-\widetilde{T}_{\ast}''(0)/(N-2)=\frac{3}{2N}+\frac{26}{N^{2}}+O(1/N^{3})$
(see Eqs. (\ref{eq_eta_1l},\ref{eq_etab_1l},\ref{eq_eta2_1l},\ref{eq_etab2_1l})):
\begin{align}
  \label{eq_eta_1l_LN}
  \eta&=\frac {3\epsilon}{2N}\left(1+\frac{52}{3N}+\mathcal
  O\left(\frac1{N^2}\right)\right)\\
\label{eq_etab_1l_LN}
  \bar\eta&=\frac \epsilon 2\left(1 +\frac{49}{N}+\mathcal
  O\left(\frac1{N^2}\right)\right)\\
  \label{eq_eta2_1l_LN}
  \eta_2&=\frac {3\epsilon}{2}\left(1+\frac{58}{3N}+\mathcal
  O\left(\frac1{N^2}\right)\right)\\
\label{eq_etab2_1l_LN}
  \bar\eta_2&=2\epsilon \left(1 +\frac{26}{N}+\mathcal
  O\left(\frac1{N^2}\right)\right) .
\end{align}

Note that a Schwartz-Soffer-like inequality is satisfied, as it
should be (see \ref{sec_one-loop_a}), by the exponents $\eta_{2}$ and
$\bar{\eta}_{2}$ (namely, $\bar{\eta}_{2}<2\eta_{2}$), but
\textit{not} by the exponents $\eta$ and $\bar{\eta}$ !
\begin{figure}[htb]
  \centering
  \includegraphics[width= \linewidth]{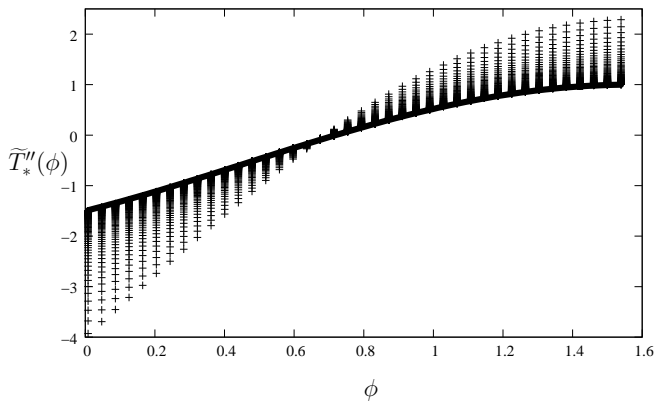}
  \caption{Variation with $N$ of the (cuspy) fixed point of the
    RA$O(N)$M in $d=4+\epsilon$ (at 1 loop):
    $\widetilde{T}_{\ast}''(\phi)$ versus $\phi$ for $0\leq\phi\leq
    \pi/2$. The curves correspond to $N=16$ (top curve in the right
    part of the figure) to $N=1500$ (bottom curve in the right part of
    the figure) by steps of $5\%$ relative increase. The thick
    curve is the analytical result for $N\to\infty$ (see
    Eq.~(\ref{eq_Sp_fp})). Note that there is a zero slope when
    $\phi=\pi/2$ and a nonzero one (\ie{}, a cusp) when $\phi\to 0^+$.
  }
  \label{fig_ra_large_n}
\end{figure}

\begin{figure}[htbp]
  \centering
  \includegraphics[width= \linewidth]{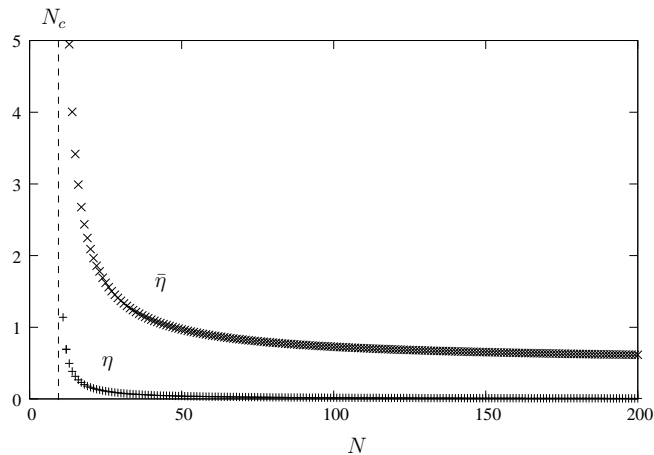}
  \caption{Exponents $\eta$ and $\bar \eta$ (divided by $\epsilon$)
  for the RA$O(N)$M at first order in $\epsilon=d-4$. Dimensional
  reduction fails for all values of $N$: $\epsilon^{-1}\eta\sim
  3/(2N)$ and $\epsilon^{-1}\bar\eta\sim
  1/2$ when $N\to\infty$. The critical value of $N$ at which both
  $\eta$ and $\bar\eta$ diverge is $N_c=9.4412...$. }
  \label{fig_eta_etab_ra_ng}
\end{figure}

We display in Figure~\ref{fig_ra_large_n} the large $N$ cuspy
fixed-point solution, Eq.~(\ref{eq_Sp_fp}), together with the
numerical solution obtained for a wide span of $N$. The convergence to
the $N\rightarrow\infty$ limit is clearly visible, as is visible the
presence for all $N$'s of a nonzero slope as $\phi\rightarrow0$, which
corresponds to a cusp. In Figure~\ref{fig_eta_etab_ra_ng} we display
the exponents $\eta$ and $\bar{\eta}$ as a function of $N$; this again
illustrates that dimensional reduction fails for all values of $N$.

\begin{figure}[htbp]
  \centering
  \includegraphics[width= \linewidth]{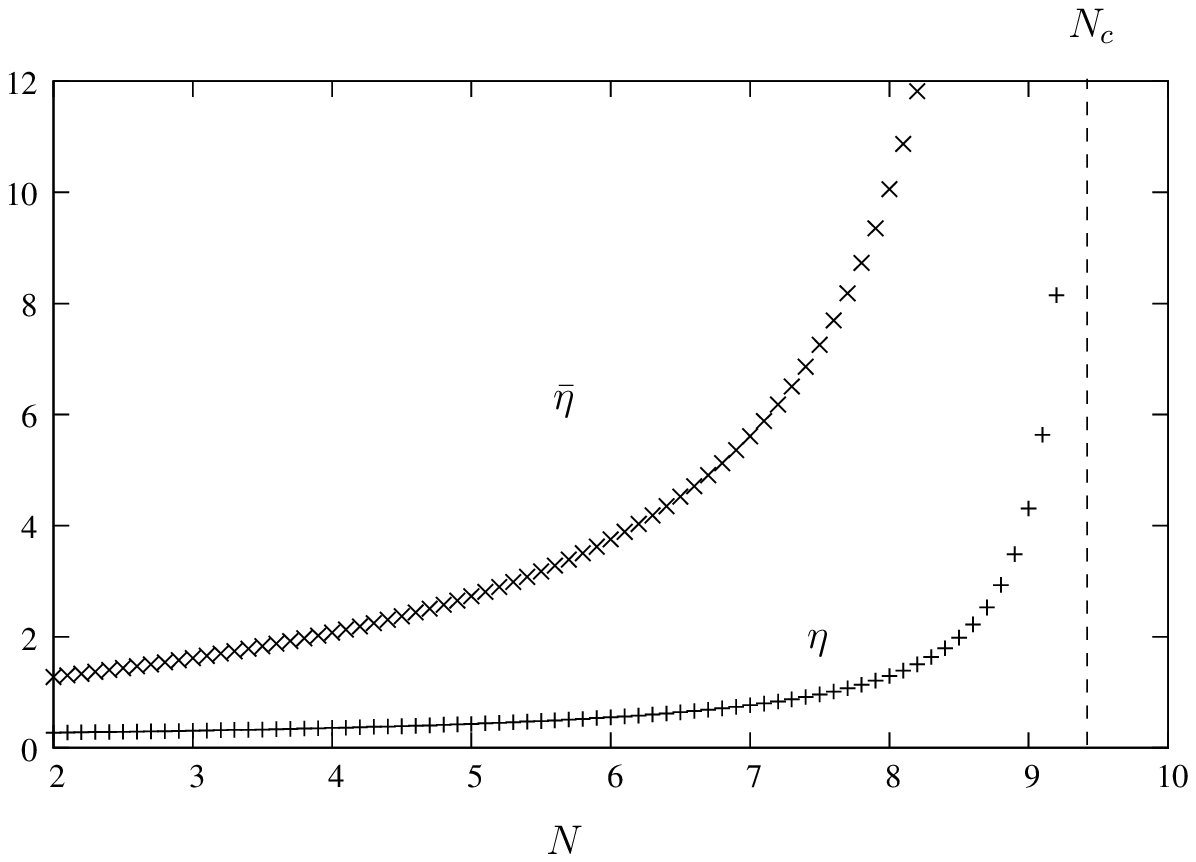}
  \caption{Exponents $\eta$ and $\etab$ (divided by $\epsilon$)
    characterizing the power-law decay of the pair correlations in the
    QLRO phase of the RA$O(N)$M for $N<N_c =9.4412...$ (first order in
    $\epsilon=d-4$). }
  \label{fig_eta_etab_ra_np}
\end{figure}

The change from ferromagnetic ordering to QLRO occurs for
$N_{c}=9.441...$, to be compared to $N_{c}=2.834...$ for the
RF$O(N)$M. As first shown by Feldman,\cite{feldman00} QLRO
may thus be present in the RA$O(N)$M near, but below, $d=4$ for $N=2,
3, ...,9$: see also Figure~\ref{fig_eta_etab_ra_np}. The (cuspy)
fixed-point solution associated with QLRO can be analytically derived
in the $N=2$ ($XY$) case. Just like in the RF$O(N)$M, the FRG
equations then reduce to those of a disordered periodic elastic
system,\cite{giamarchi94,giamarchi95,giamarchi98} the only difference with the RF case being a
simple rescaling of the solution accounting for the difference in the
periodicity, from $2\pi$ to $\pi$. This yelds the critical exponents
$\eta=|\epsilon|\pi^2/36$ and $\bar{\eta}=|\epsilon|(1+\pi^2/36)$.

\section{Derivation of the FRG beta functions to 2 loops}
\label{sec_two_loops}

In this section we describe in detail the calculation of the beta
functions and of the critical exponents at two loops.

\subsection{Diagrammatic representation}

As explained above (see section \ref{sec_pert_renorm}) the calculation
is based on an expansion of the effective action in powers of $R$. The
terms of this expansion are given by all amputated 1-particle
irreducible Feynman diagrams. In order to determine all the
counterterms, we need to compute the 1-replica 2-point proper vertex
and the 2-replica effective action with no derivatives (uniform
fields).  The associated diagrams are obtained by connecting the
different vertices of the theory with the free propagator given in
Eq.~(\ref{eq_propag}).

\begin{figure}[htbp]
   \centering \includegraphics[width=.25 \linewidth, bb=145 605 210
   670]{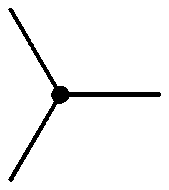} \includegraphics[width=.25 \linewidth, bb=145 605
   210 670]{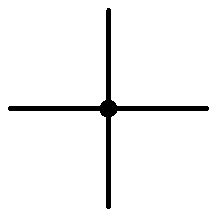} \includegraphics[width=.25 \linewidth, bb=145
   605 210 670]{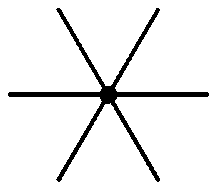}
   \caption{Vertices with 3 and 4 legs obtained from $\mathcal S_1$, \ie{},  with 1 replica.}
   \label{fig_vert_1_replica}
\end{figure}

\begin{figure}[htbp]
   \centering \includegraphics[width=.24 \linewidth, bb=145 605 210
   670]{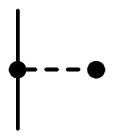} \includegraphics[width=.24 \linewidth, bb=145
   605 210 670]{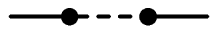} \includegraphics[width=.24 \linewidth,
   bb=145 605 210 670]{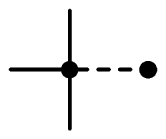} \includegraphics[width=.24
   \linewidth, bb=145 605 210 670]{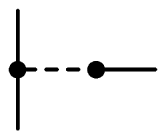}
   \caption{Vertices with 2 and 3 legs obtained from $\mathcal S_2$, \ie{},  with 2 replicas.}
   \label{fig_vert_2_replica}
\end{figure}

The free propagator is represented by a line. The vertices are
obtained by deriving either $\mathcal S_1$ (Eq.~\ref{eq_S1_re}) or
$\mathcal S_2$ (Eq.~\ref{eq_S2_re}). In the former case they are
represented by lines emerging from a single circle (see
Fig.~\ref{fig_vert_1_replica}) and in the latter
by lines emerging from two circles (corresponding to the
two replicas of $\mathcal S_2$) connected by a dashed line (see
Fig.~\ref{fig_vert_2_replica}).

\begin{figure}[htbp]
  \centering \includegraphics[width=.19 \linewidth, bb=145 605 265
  670]{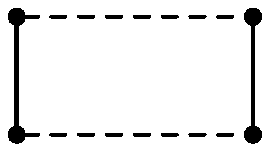} \includegraphics[width=.19 \linewidth, bb=145 605
  265 670]{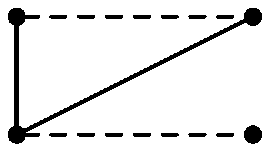}
\caption{1-loop diagrams for the  2-replica effective action.} 
\label{fig_1loop_2rep}
\end{figure}

\begin{figure}[htbp]
 \centering \includegraphics[width=.19 \linewidth, bb=145 605 265
  670]{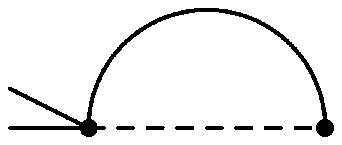} \includegraphics[width=.19 \linewidth, bb=145 605
  265 670]{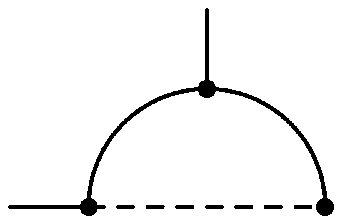} \includegraphics[width=.19 \linewidth, bb=145
  605 265 670]{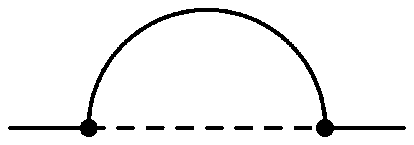} \includegraphics[width=.19 \linewidth,
  bb=145 605 265 670]{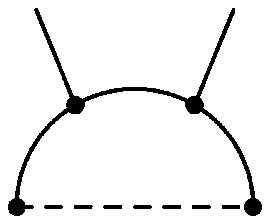} \includegraphics[width=.19
  \linewidth, bb=145 605 265 670]{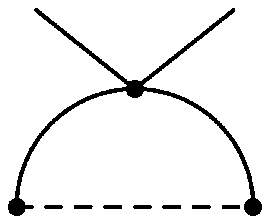}
\caption{1-loop diagrams for the 1-replica 2-point proper vertex.} 
 \label{fig_1loop_1rep}
\end{figure}

As can be seen in Eqs.~(\ref{eq_S1},\ref{eq_S2}) the 1- and 2-replica
parts of the action have a factor $T^{-1}$ and
$T^{-2}$ respectively, so that the various diagrams do not come with the same
power of the temperature. Anticipating that the fixed point of
interest to us is at zero temperature, we compute here only the
diagrams of lowest order in $T$, \ie{}, those proportional to $1/T$ for the
1-replica effective action and to $1/T^2$ for the 2-replica effective
action.  It is easy to check that the diagrams of lowest order in
temperature with $n_1$ 1-replica vertices and $n_2$ 2-replica vertices
have $n_1+2n_2-2$ propagators. Similarly, for the 2-replica effective
action, the diagrams of lowest order in temperature have $n_1+2n_2-1$
propagators.  Given these constraints, one can draw all the diagrams
and check that an expansion in powers of $R$ corresponds to an
expansion in increasing number of loops.  The 1-loop diagrams are
displayed in Figure~\ref{fig_1loop_2rep} and Figure~\ref{fig_1loop_1rep}.


The 2-loop diagrams are given in Appendix \ref{sec_diagrams}. The
integrals involved in the 2-loop calculations have been evaluated in the dimensional regularization scheme by
using the procedure described in Appendix~\ref{sec_integrals}.

The 1-replica 2-point proper vertex can be formally expressed as
\begin{equation}
  \label{eq_1replica_2loops}
\begin{split}
  \frac{\delta^2\Gamma_1[\vect\pi]}{\delta\pi^i(q)
    \delta\pi^j(q)}= {Z_\Pi}&\Bigg(\frac 1T\left(q^2-\frac H\Sigma\right)
    \left (\delta_{ij}+\frac{\Pi_i
    \Pi_j}{\Sigma^2}\right)\\&+\text{1-loop} +\text{2-loops}\Bigg).
\end{split}
\end{equation}
with $\Sigma=1-\vect\Pi^{2}$. The first term in the parentheses
corresponds to the tree diagram and is given by the inverse of the
free propagator (see Eq.~(\ref{eq_propag})).

The same can be done for the 2-replica effective action, which
formally gives
\begin{equation}
  \label{eq_2replica_2loops}
\begin{split}
 \Gamma_2[\vect\Pi_a,\vect\Pi_b]=\frac 1{T^2} 
  R_0(\vect\Pi_a \cdot
  \vect\Pi_b&+\Sigma_a\Sigma_b)\\&+\text{1-loop} +\text{2-loops}.
\end{split}
\end{equation}

We next replace the bare quantities by the renormalized ones,
following Eqs.~(\ref{eq_renormalized}), in the two previous
expressions and reexpand in powers of $R$. One must then choose the
couterterms such that the expressions are finite, \ie{}, such that all
terms in $1/\epsilon$ and $1/\epsilon^{2}$ vanish.\footnote{Note that,
  in the 2-loop calculation, one must keep the 1-loop contributions
  that are finite (of order $O(\epsilon^0)$) because these terms, when
  the replacements defined in Eqs.~(\ref{eq_renormalized}) are made,
  are multiplied by the 1-loop counterterms in $1/\epsilon$ and
  therefore contribute to the singular part.}

Once the counterterms are known, the beta functions can be calculated
as derivatives of the renormalized quantities with respect to the
scale $\mu$ at fixed bare quantities (see
Eqs.~(\ref{eq_zeta_pi},\ref{eq_zeta_T},\ref{eq_betaR})). In order to
perform the derivative in Eq.~(\ref{eq_betaR}), we write the
counterterm for $R$ as:
\begin{equation}
  \label{eq_ct_r_exp}
  Z_R=R+\delta_1[R,R]+\delta_2[R,R,R]
\end{equation}
where $\delta_1[R,R]$ and $\delta_2[R,R,R]$ are the 1-loop and 2-loop
contributions, respectively; the former is a quadratic functional of
$R$ and is proportional to $1/\epsilon$, and the latter is cubic in
$R$ and contains terms in $1/\epsilon^2$ and $1/\epsilon$. We then
invert Eq.~(\ref{eq_renorm_R}) with $Z_R$ given by
Eq.~(\ref{eq_ct_r_exp}), so that the renormalized function $R$ can be
expressed as a functional of the bare function $R_0$ as
\begin{equation}
  \label{eq_R_renorm_bare}
  \begin{split}
    R=\mu^{\epsilon} R_0-& \mu^{2\epsilon}
        \delta_1[R_0,R_0]-\mu^{3\epsilon}
        \big(\delta_2[R_0,R_0,R_0]\\ &-
        2\delta_1[\delta_1[R_0,R_0],R_0]\big)+O(R_0^4).
  \end{split}
\end{equation}

The term on the last line corresponds to a repeated 1-loop term, in which the
first argument of the functional $\delta_1$ is replaced by $\delta_1[R_0,R_0]$. This
gives a cubic contribution in $R_0$. The flow equation for
$R$ is then given by Eq.~(\ref{eq_betaR}) with the following beta function:
\begin{equation}
  \label{eq_beta_ct}
  \begin{split}
    {\beta_R}=-\epsilon\big(R-\delta_1[R,R]-&2\delta_2[R,R,R]+2
\delta_1[\delta_1[R,R],R]\big).
  \end{split}
\end{equation}

In the above expression, the terms in $1/\epsilon$ appearing in
$\delta_1$, $\delta_2$, and in the repeated 1-loop term are multiplied by
$\epsilon$ and therefore give finite contributions. The last two terms also have
contributions in $1/\epsilon^2$, but they exactly cancel each other (as can be
explicitly checked by using Eqs.~(\ref{eq_ctR1}) and (\ref{eq_ctR2}) below).

\subsection{Analytic piece of the beta functions}
\label{sec_analytic}
As discussed in section \ref{sec_one-loop}, even when starting from an
analytic initial condition, integration of the 1-loop flow
equation generates a nonanalyticity in the renormalized disorder function $R(z)$. It is therefore necessary to consider the flow equations for
nonanalytic functions $R(z)$, and in particular to determine the
contributions of such nonanalyticities to the beta functions. At the
1-loop order, there are no such contributions. However, they
 appear at 2 loops. For clarity's sake, we first
describe the calculation of the flow equations for analytic functions,
and consider the contributions due to nonanaliticities in the next section.

The analysis described in the previous subsection and in Appendices B and C leads to the following expressions for the counterterms:
\begin{widetext}
\begin{align}
  \label{eq_Zpi}
  Z_\Pi^{\text r}=1+\frac C\epsilon(N-1)&R'(1)\left(1+
    \frac{C}{2\epsilon}R'(1)(2N-3)\right)\\
  \label{eq_ZT}
  \begin{split}
    Z_T^{\text r}=1+\frac
    C\epsilon(N-2)&R'(1)\bigg(1+\frac{C}{\epsilon}R'(1)(N-2+\frac\epsilon
    2)\bigg)
  \end{split}
\end{align}
  \begin{equation}
  \label{eq_ctR1}
  \begin{split}
   \frac{2 \epsilon} {C} \delta_1[R,R]=& \left(z^2+N-2\right)
   R'(z)^2+2 z \left((1-N) R'(1)+\left(z^2-1\right) R''(z)\right)
   R'(z)+4 (N-2) R(z) R'(1)\\&+\left(z^2-1\right) R''(z)
   \left(\left(z^2-1\right) R''(z)-2 R'(1)\right)
  \end{split}
\end{equation}
\begin{equation}
  \label{eq_ctR2}
  \begin{split}
    &\frac{4 \epsilon^2}{C^2}\delta_2^{\text r}[R,R,R]=(\epsilon +2) R''(z)
    R'''(z)^2 (z^2-1)^4+2 R''(z)^2 (3 z (\epsilon +4)
    R'''(z)+(z^2-1) R''''(z)) (z^2-1)^3+\\&((9 \epsilon +32)
    z^2+N (\epsilon +2)-2 (\epsilon +6)) R''(z)^3 (z^2-1)^2+R'(z) (z
    (15 (\epsilon +4) z^2+N (3 \epsilon +10)-2 (6 \epsilon +25))
    R''(z)^2+\\&2 (z^2-1) ((4 (\epsilon +5) z^2+2 N-\epsilon -6)
    R'''(z)+2 z (z^2-1) R''''(z)) R''(z)+z (z^2-1)^2 (\epsilon
    +2) R'''(z)^2) (z^2-1)+\\&z ((\epsilon +4) z^2+2 N-\epsilon -6)
    R'(z)^3+4 (N-2) (3 N+\epsilon -6) R(z) R'(1)^2+2 R'(1)^2 ((-4
    N^2+13 N-9) z R'(z)+\\&(N^2 z^2+9 z^2-3 N (z^2-1)-10)
    R''(z)+(z^2-1) (2 (N+1) z R'''(z)+(z^2-1) R''''(z)))+R'(z)^2
    (((7 \epsilon +32) z^4\\&-12 (\epsilon +4) z^2+2 N^2+5 \epsilon +2
    N (z^2-1) (\epsilon +5)+14) R''(z)+2 z (z^2-1) (((\epsilon +8)
    z^2+2 N-\epsilon -6) R'''(z)+\\&z (z^2-1) R''''(z)))-R'(1)
    ((\epsilon +2) R'''(z)^2 (z^2-1)^3+2 R''(z) (z (2 N+3 \epsilon
    +14) R'''(z)+2 (z^2-1) R''''(z)) \\&(z^2-1)^2+((7 \epsilon
    +34) z^2-4 (\epsilon +6)+N ((\epsilon +6) z^2+2 (\epsilon +2)))
    R''(z)^2 (z^2-1)+(-4 N^2+((\epsilon -2) z^2-\epsilon +\\&18) N-z^2
    (\epsilon -10)+\epsilon -22) R'(z)^2+2 R'(z) (z (2 N^2+(z^2-1)
    (\epsilon +2) N-\epsilon +z^2 (\epsilon +18)-20) R''(z)+\\&(z^2-1)
    (((\epsilon +10) z^2+2 N (z^2+1)-\epsilon -6) R'''(z)+2 z
    (z^2-1) R''''(z))))
      \end{split}
\end{equation}
\end{widetext}
where the superscript $\text r$ stands for `regular' \ie{}, analytic),
$C^{-1}=8\pi^2$ and $z=\vect\pi_a\cdot\vect\pi_b+\sqrt{1-\vect\pi_a^2}\sqrt{1-\vect\pi_b^2}$.

Note that the expressions for $Z_\Pi$ and $Z_T$ correspond to those
obtained at two loops in the nonlinear sigma model for the
ferromagnetic-paramagnetic transition of the $O(N)$ model with no randomness if
one replaces $R'(1)$ by the temperature, $C$ by $1/2\pi$, and
$\epsilon$ by $d-2$.\cite{brezin76} Moreover the counterterm for $R'(1)$ (assuming
here that the function $R(z)$ is analytic) reads
\begin{align}
  \begin{split}
\frac 1{R'(1)}  \partial_z(\delta_1+&\delta_2^{\text r})|_{z=1}=\frac
C\epsilon R'(1)(N-2)\\&\left(1 + \frac
C\epsilon R'(1)\left(N-2+\frac \epsilon 2\right)\right),
  \end{split}
\end{align}
which again coincides with the couterterm found for the temperature in the pure system.\cite{brezin76} This equivalence between the pure
model near $d=2$ and the disordered one near $d=4$ is the expression of the dimensional-reduction property.

The regular part of the beta function for $R(z)$ can then be calculated from Eqs.~(\ref{eq_beta_ct},\ref{eq_ctR1},\ref{eq_ctR2}), which gives
\begin{widetext}
\begin{equation}
  \label{eq_beta_analytic}
  \begin{split}
    \beta_R^{\text r}&(z)=-\epsilon R(z) +\frac{C}{2} \Big[(N-2+z^2)
    R'(z)^2-2 z ((1-z^2) R''(z)+(N-1) R'(1)) R'(z)+4 (N-2) R(z)
    R'(1)+\\&(1-z^2) R''(z) ((1-z^2) R''(z)+2 R'(1))\Big]
    +\frac{C^2}{2}\Big[ \left(1-z^2\right) \left(\left(1-z^2\right)
      R''(z)+R'(1)-z R'(z)\right)\\& \left(\left(1-z^2\right)
      R'''(z)-3 z R''(z)-R'(z)\right)^2+(N-2)
    \Big(\left(1-z^2\right)^2 R''(z)^3-\left(1-z^2\right) \left(3 z
        R'(z)-\left(z^2+2\right) R'(1)\right) R''(z)^2\\&-2
      \left(1-z^2\right) R'(z) \left(R'(z)-z R'(1)\right)
      R''(z)+\left(1-z^2\right) R'(1) R'(z)^2+4 R(z)
      R'(1)^2\Big)\Big].
      \end{split}
\end{equation}
\end{widetext}

We next consider the derivation of the critical exponents. The determination of
$\zeta_\Pi$ and $\zeta_T$ simplifies if one uses the fact that $Z_\Pi$
and $Z_T$ depend on $R(z)$ only through $R'(1)$. Eqs. (\ref{eq_zeta_pi},\ref{eq_zeta_T}) become
\begin{equation}
  \zeta_A^{\text r}=\mu\partial_\mu\log Z_A^{\text r}=- 
\beta_{R'(1)}\partial_{R'(1)}\log Z_A^{\text r}
\end{equation}
with $A$ being $\Pi$ or $T$ and the derivatives being taken at fixed bare quantities. We then get

\begin{align}
  \zeta_\Pi^{\text r}&=(N-1)CR'(1)+\mathcal O(R^3)\\
  \zeta_T^{\text r}&=(N-2)C R'(1)(1+C R'(1))+\mathcal O(R^3).
\end{align}

The critical exponents $\eta$ and $\bar \eta$ can now be evaluated by
making use of 
Eqs.~(\ref{eq_eta},\ref{eq_etab}):
\begin{align}
  \eta^{\text r}=&CR_\ast'(1)(1-(N-2)CR_\ast'(1))\\
  \bar \eta^{\text r}=&-\epsilon+(N-1)CR_\ast'(1).
\end{align}

As we shall explicitly show in section V, the exponents defined under
the assumption of an analytic fixed-point solution $R_{\ast}(z)$ are
equal to their dimensional-reduction value.

\subsection{'Anomalous' contributions}
\label{sec_nonanalytic}

We have seen in section III that, even with an analytic initial
condition for $R_{t=0}(z)$, the 1-loop RG flow equation generates a
nonanaliticity in the renormalized disorder function $R(z)$.  The
strongest nonanaliticity is obtained in the RF$O(N)$M for $N<18$ and
in the RA$O(N)$M for all values of $N$ in the form

\begin{equation}
  \label{eq_cusp}
  R(z)= R(1)+R'(1)(z-1) -\frac a3
\left(2(1-z)\right)^{\frac 32}+\dots
\end{equation}
when $z$ approaches 1 (from below); $a$ will be used in the following
to quantify the strength of the singularity. Dimensional reduction is
recovered under the assumption that the function $R(z)$ is analytic
(see above), but if $a\neq 0$ it is no longer valid.

Alternatively, the renormalized disorder function can be parametrized
in terms of the angle $\phi$ between the two replicas instead of the
scalar product $z$ ($z=\cos\phi$). The expansion in
Eq.~(\ref{eq_cusp}) then translates into a small $\phi$ expansion,
\begin{equation}
  \label{eq_cusp_theta}
  R(\phi)\equiv R(z=\cos\phi)= R(1)-R'(1)\frac {\phi^2}2 -\frac {a
|\phi|^3}3+\dots
\end{equation}
where the nonanalyticity appears as a discontinuity in the third derivative
of $R(\phi)$ in $\phi=0$.

One can easily convince oneself that the nonanalytic term in $a$ can
explicitly appear in $\beta_R(z)$. Consider the repeated 1-loop term
(last term in Eq.~(\ref{eq_beta_ct})). Since $\delta_1[R,R]$ has an
explicit dependence on $R'(1)$, one has to compute $\partial_z
\delta_1|_{z=1}$ which, when the nonanalyticity of $R(z)$ is taken
into account (see Eq.~(\ref{eq_cusp})), takes the form
\begin{equation*}
\partial_z\delta_1|_{z=1}=\frac C\epsilon\left
  (R'(1)^2(N-2)-a^2(N+2)\right).
\end{equation*}
Replacing then $R'(1)$ by Eq.~\ref{eq_cusp} yields an explicit
dependence of $\beta_R$ on $a$.

Actually, the 2-loop, 2-replica diagrams also give contributions in
$a^2$: look for instance at the sixth diagram of
Figure~\ref{fig_diag_2loops} in the Appendix~\ref{sec_diagrams}. Note
that the two replicas of the vertices on the left and on the right of
the diagram are actually connected via propagators. These vertices are
therefore to be computed for identical replicas. In order to take into
account the nonanalytic part of this diagram, the 2-replica vertices
are evaluated for two slightly different replicas $\vect S_{a}$ and
$\vect S_{a'}$ with

\begin{equation}
  \label{eq_limit}
  \vect S_{a'}=\frac{\vect S_{a}+\alpha\vect
  \Delta}{\sqrt{1+\alpha^2}}.
\end{equation}
Here, $\alpha$ is a small parameter that must be taken to zero at the
end of the calculation and $\vect \Delta$ is a unit vector orthogonal
to $\vect S_a$ that gives the direction in which $\vect S_{a'}$
approaches $\vect S_a$. The dependence of the diagram on $\vect
\Delta$ appears only through the scalar product $\vect \Delta\cdot
\vect S_b$ whose absolute value varies between $0$ (when $\vect \Delta$
and $\vect S_b$ are orthogonal) and $\sqrt{1-z^{2}}$ (when $\vect \Delta$,
$\vect S_a$ and $\vect S_b$ are in the same plane). We therefore
write

\begin{equation}
  \label{eq_limit_mup}
\vect \Delta\cdot \vect S_b=\gamma\sqrt{1-z^{2}}  
\end{equation}
with $\gamma$ varying between $-1$ and $1$.

There are six 2-replica diagrams giving nonanalytic contributions in
$a^2$: diagrams 5, 6, 11, 12, 13 and 14 of
Figure~\ref{fig_diag_2loops}. These diagrams (and some others) also
have contributions linear in $a$, but we discard them for symmetry
reasons. Indeed, $a$ corresponds to the third derivative of $R(\phi)$
(see Eq.(\ref{eq_cusp_theta})) which changes sign under the operation
$\phi\to-\phi$. On the other hand, the disorder function itself is
unchanged in the same operation; as a result, linear contributions in
$a$ must vanish from all physical quantities.

The situation is even more complex when considering the 2-loop
diagrams for the 1-replica 2-point function. In this case a generic
diagram has a singular limit $\alpha\to 0$. Indeed an expansion in
powers of $\alpha$ around $0$ gives terms in $a^2 \alpha^{-2}$, $a^2
\alpha^{-1}$ and $a^2 \alpha^{0}$. On top of this, there is a strong
dependence of the result on the way the regularization is performed.
Look for instance at the first diagram of Figure~\ref{fig_diag_B}. We
can decide to attribute the field $\vect S_a$ to the replica on the
left of the diagram and the field $\vect S_{a'}$ to the replica on
the right. It remains to choose whether the propagator (which is diagonal in replica indices) on the top of
the diagram is associated with $\vect S_a$ or $\vect
S_{a'}$. The calculation shows that the two choices lead to
different results !

Such ambiguities are already present in the 2-loop FRG treatment of
disordered elastic systems and Le Doussal, Wiese and coworkers have
given a detailed and well-argumented analysis of the way to handle
these ambiguities.\cite{ledoussal04} Here, we have extended their
procedure and used the following set of rules:

\begin{enumerate}
\item The nonanalytic parts of the diagrams, \ie{}, those proportional to $a^{2}$, come with an \textit{a priori} unknown weight.
\item Within a single diagram, the parts in $a^2 \alpha^{-2}$, $a^2
  \alpha^{-1}$ and $a^2 \alpha^{0}$ come with independent (\textit{a priori} unknown) weights.
\item For diagrams that are ambiguous in the sense that different
  regularization schemes lead to different results, we have introduced
  additional weighting factors such that all possible results can be
  reproduced by appropriately choosing these extra weighting factors.

\end{enumerate}

As discussed in Ref.~\onlinecite{ledoussal04}, the fact that pieces
of the 2-loop diagrams come with \textit{a priori} unknown weighting
factors is due to the intrinsic ambiguity that occurs at $T=0$ when
the function $R$ entering into the vertices is nonanalytic in $z=1$
(or $\phi=0$). Consider for instance the third derivative of $R(\phi)$
around $\phi=0$ (see Eq.~\ref{eq_cusp_theta}). Its sign depends on
whether $\phi \rightarrow 0^{+}$ or $\phi \rightarrow 0^{-}$ and
because of the discontinuity its value in exactly $\phi=0$ is left
undetermined. Vertices that contain this derivative evaluated exactly
in $\phi=0$ have thus a contribution that come with an undetermined
weight. Additional constraints must be used to fix the values of the
weighting factors (or at least enough relations between these
factors). This is precisely what is done by requiring that the
physical quantities be finite.

Under these hypotheses, the calculation proceeds in a straightforward
way. We observe that it is possible to choose the nonanalytic part of
the counterterms and to fix all the weighting factors such that the
1-replica 2-point proper vertex and the 2-replica effective action
(see Eqs.(\ref{eq_1replica_2loops}) and (\ref{eq_2replica_2loops}))
are finite. This procedure leads to a \textit{unique} form for the
counterterms (for a given parameter $\gamma$, see
Eq.~(\ref{eq_limit_mup})).  On top of the analytic parts already
computed (see Eqs.(\ref{eq_Zpi}), (\ref{eq_Zpi}) and (\ref{eq_ctR2})),
one now must add the 'anomalous' contributions, so that

\begin{equation}
  \label{eq_Zpi_nona}
  Z_\Pi=Z_\Pi^{\text r}-\frac{C^2}{\epsilon^2}a^2\left( \frac
  {(N-1)(N+2)}2+\epsilon\frac{3N-2}4 \right)
\end{equation}
\begin{equation}
  \label{eq_ZT_nona}
  Z_T=Z_T^{\text r}-\frac{C^2}{\epsilon^2}a^2\left( \frac
  {N^2-4}2+\epsilon\frac{N-2}2 \right)
\end{equation}
\begin{equation}
  \label{eq_ctR2_nana}
  \begin{split}
    \delta_2[R,&R,R]=\delta_2^{\text r}[R,R,R]+\frac
    {C^2a^2}{\epsilon^2}\Big\{\\&\left(4-N^2-
    (N-2)\epsilon\right)R(z)+\\&\frac z
    4\left(2(N-1)(N+2)+\epsilon(3N-2)\right)R'(z)\\&-\frac{1-z^2}4
    \left(2N+4+(2+\gamma^2N) \epsilon\right)R''(z)\Big\}.
  \end{split}
\end{equation}
The presence of 'anomalous' contributions in the counterterms induce new terms in the beta function for $R(z)$, which now reads
\begin{equation}
  \label{eq_beta_nona}
  \begin{split}
    \beta_R&(z)=\beta_R^{\text r}(z)-\dfrac{C^2 a^2}{2}\Big\{ 4(N-2)R(z)-\\&(3N-2)zR'(z)+(1-z^2)(\gamma^2N+2)R''(z)\Big\},
      \end{split}
\end{equation}
where, we recall, $a$ is defined through Eq.~(\ref{eq_cusp}). Note here
that there is still an explicit dependence on $\gamma^2$, which
encodes how one takes the limit $\vect S_{a'} \to \vect S_{a}$ (see
Eq.~(\ref{eq_limit_mup})). There is however a preferred value for
$\gamma^2$. The simplest way to see this is to compute $\beta_R(z)$
with a nonanalytic function $R$ of the form given in
Eq.~(\ref{eq_cusp}). There is a term proportional to $a^3
(1-\gamma^2)\sqrt{1-z}$. If we choose $\gamma^2\neq 1$, the flow
equation generates a {\em supercusp}, \ie{} a stronger nonanalyticity,
$R(z)\sim(1-z)^{1/2}$, than the one initially considered; this
supercusp would itself generate an even stronger nonanalyticity and
the theory would not be renormalizable at 2-loop level. On the other
hand, $\gamma^2=1$ ensures that the procedure is consistent. The value
$\gamma^2=1$ also appears in another context: the repeated 1-loop term
(last term of Eq.~(\ref{eq_beta_ct})) can be interpreted as the set of
2-loops diagrams obtained by replacing in the 1-loop diagrams (see
Figure \ref{fig_1loop_2rep}) one of the 2-replica vertices by the
1-loop diagrams with two external legs. If we compute these 2-loop
diagrams with their weighting factor according to the above procedure,
we get an expression that is consistent with the repeated term only if
we choose $\gamma^2=1$. In the following, we thus fix $\gamma^2$ to
this value.

It is worth stressing that
Eqs.~(\ref{eq_beta_analytic},\ref{eq_beta_nona}) (with $\gamma^2=1$)
exactly reduce to the 2-loop FRG equation for periodic disordered
elastic systems when $N=2$.\cite{ledoussal04} This is more easily
checked by switching to the angle variable $\phi$.

In Refs.~(\onlinecite{tissier06,ledoussal06}) the renormalization
constant for the temperature $Z_{T}$ had not been derived at two
loops, so that an incompletely determined version of the beta function
of Eq.~\ref{eq_beta_nona} was given. In the notations of
Ref.~\onlinecite{tissier06} the unknown parameter $K$ is now fixed to
$K=1/2$ whereas in the notations of Ref.~\onlinecite{ledoussal06}, the
unknown parameter is now fixed to $\gamma_a=1/4$.\footnote{This value
  corresponds to the lower bound given in footnotes 27 and 28 of
  Ref.~\onlinecite{ledoussal06}.}

Finally, the expressions of the critical exponents have to be modified
in order to take into account the nonanalytic contributions. One
obtains
\begin{equation}
  \label{eq_eta_nona}
  \eta=\eta^{\text r}-\frac{C^2a_{\ast}^2}2(N+2)
\end{equation}
\begin{equation}
  \label{eq_etab_nona}
  \bar\eta=\bar\eta^{\text r}-\frac{C^2a_{\ast}^2}2(3N-2),
\end{equation}
where $a_{\ast}$ is the fixed-point value for the parameter $a$.

\section{Discussion of the fixed points at the 2-loop level and
  conclusion}
\label{sec_fp_2-loop}

Going from the 1-loop to the 2-loop order does not significantly alter
the general behavior of the RF and RA models in $d=4+\epsilon$. As we
shall see, it allows nonetheless to show that no
Kosterlitz-Thouless-like transition occurs at the special point
($N_{c},d=4$) and that in the vicinity of this point, for $N<N_{c}$
and $d<4$, a once unstable fixed point appears, which describes the
transition from the QLRO phase to the paramagnetic one. The picture is
now fully compatible with that found in our NP-FRG approach and
summarized in Figure~\ref{fig_phase_diag_np}.

As for the 1-loop level, it is worthwhile to start by considering the
beta functions for the first two derivatives of the renormalized
disorder correlator $R(z)=(\epsilon/C)\widetilde{R}(z)$ evaluated in
$z=1$, assuming that the second derivative is well defined at that
point. Writing the beta functions as $\beta=\beta_{1}+\beta_{2}$, we
only give the expressions for the 2-loop contributions $\beta_{2}$, the
1-loop terms $\beta_{1}$ being given in Eqs.~(\ref{eq_beta_Rp1},\ref{eq_beta_Rs1}):

\begin{equation}
  \label{eq_2l_Rp}
  \epsilon^{-2}\beta_{\widetilde R'(1) ,2}=(N-2)\widetilde R'(1)^3
\end{equation}
\begin{equation}
  \label{eq_2l_Rs}
  \begin{split}
    \epsilon^{-2}&\beta_{\widetilde R''(1) ,2}=2(5N+17)\widetilde
    R''(1)^3+\\&6(N+7)\widetilde R'(1)\widetilde
    R''(1)^2-6(N-5)\widetilde R'(1)^2\widetilde R''(1)\\&-(N-4)\widetilde
    R'(1)^3 .
  \end{split}
\end{equation}

One can immediately see that the fixed-point solution
$\widetilde{R}_{\ast}'(1)=\frac{1}{N-2}(1-\frac{\epsilon}{N-2})$ and its associated
(positive) eigenvalue $\Lambda_{1}=\epsilon(1+\frac{\epsilon}{N-2})$ lead to
dimensional reduction,
$\eta=\bar{\eta}=\eta_{DR}=\frac{\epsilon}{(N-2)}(1+\frac{N-1}{N-2}\epsilon)$ (after using Eqs. (60,61)) and
$\nu=\nu_{DR}=\frac{1}{\epsilon}-\frac{1}{N-2}$. When $\widetilde{R}'(1)$
is chosen equal to its fixed-point value, the beta function for the
second derivative becomes a cubic polynomial in $\widetilde{R}''(1)$ which
has zeros only if the associated discriminant is negative,
\ie{}, if $N\geq18-\frac{49}{5}\epsilon$.\cite{sakamoto06,ledoussal06}

Following the same lines as for the 1-loop order, one can check that
the RF$O(N)$M at 2 loops has a critical fixed point with a
'subcusp' when $N>N_{DR}=18-\frac{49}{5}\epsilon$, but that the
critical fixed point of the RA$O(N)$M has always a cusp, implying
$N_{DR}=\infty$ (see Appendix~\ref{sec_ra_large_n}).

We next concentrate on the region $N\leq N_{c}$ and $d\leq 4$. A first
result is that the beta function for $R'(z)$ (unscaled by $\epsilon$)
in $d=4$ does not vanish for arbitrary values of the disorder strength
$R_{\ast}'(1)$. When scaling out the disorder strength to define
$r'(z)=R'(z)/R_{\ast}'(1)$, the beta function for $\epsilon=0$ can be
expressed as
$\beta_{r'}(z)=\beta_{1}[r',r']+R_{\ast}'(1)\beta_{2}[r',r',r']$. $\beta_{r'}(z)$
then vanishes independently of the value of $R_{\ast}'(1)$ if and only
if $\beta_{1}[r',r']=0$ and $\beta_{2}[r',r',r']=0$ have the same
solution $r_{\ast}'(z)$. It is straightforward to check that inserting
the solution of the 1-loop equation $\beta_{1}=0$ into the 2-loop
equation does not make the latter vanish identically. The consequence
is that no Kosterlitz-Thouless transition can exist in ($N_{c},d=4$)
since, even perturbatively, no line of fixed points can be
found. Actually, the only fixed point at 2-loop level for $N=N_{c}$
and $d=4$ is the trivial one with $R_{\ast}'(1)=0$.

For $N\lesssim N_{c}$ and $d\lesssim 4$, consideration of the 2-loop
order brings in a new phenomenon. An additional, once unstable, fixed
point appears, which describes the transition from the QLRO phase to
the paramagnetic one. This new fixed point is found perturbatively in
$\epsilon$ (which is now negative) and $N-N_{c}$. More precisely, as
shown by Le Doussal and Wiese,\cite{ledoussal06} it can be obtained
within a double expansion in $\sqrt{\vert\epsilon\vert}$ and
$N-N_{c}$. For any given value of $N\lesssim N_{c}$, the critical
fixed point and the QLRO fixed point get closer as
$\vert\epsilon\vert$ increases and they merge for a value
$\epsilon_{lc}(N)$. This latter corresponds to the lower critical
dimension $d_{lc}=4+\epsilon_{lc}$ of QLRO. With the full 2-loop
results given above, Eqs.~(\ref{eq_beta_analytic},\ref{eq_beta_nona}), one finds
\begin{align}
  \label{eq_dlc-rf}
  d_{\text{lc}}(\text{RF})=4- 0.14 (N-N_c)^2+\mathcal O((N-N_c)^3)\\
  \label{eq_dlc-ra}
  d_{\text{lc}}(\text{RA})=4- 0.002 (N-N_c)^2+\mathcal O((N-N_c)^3)
\end{align}

where $N_c=2.8347...$ for the RF model and $9.4412...$ for the RA
model. For what it is worth, directly plugging $N=2$ in the above
expressions gives the following estimates for the lower critical
dimension of the (QLRO) Bragg glass phase in the $XY$ model:
$d_{\text{lc}}(\text{RF})\simeq 3.9, d_{\text{lc}}(\text{RA})\simeq
3.9$. (Our NP-FRG theory predicts $d_{\text{lc}}(\text{RF})\simeq
3.8$,\cite{tissier06} so that in all cases no Bragg glass is found for
$N=2$ in $d=3$.)

\begin{figure}[htb]
  \centering
  \includegraphics[width= \linewidth]{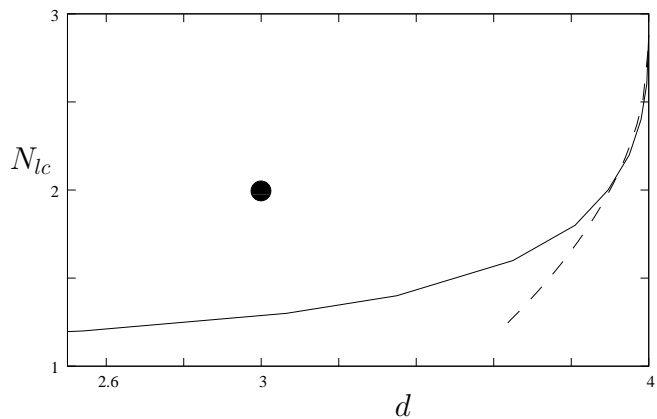}
  \caption{Comparison between the results of the 2-loop perturbative
    FRG (dashed line) and of the NP-FRG (continuous line) near $d=4$
    for the QLRO lower critical dimension $N_{lc}(d)$ in the case of
    the RF$O(N)$M. The black circle denotes the physical case of the
    XY model in $d=3$, a case which is clearly below its lower
    critical dimension.}
  \label{fig_nc_comp_qlro}
\end{figure}

Note that the fact that no Kosterlitz-Thouless transition takes place
in $N=N_{c}$ and $d=4$ is connected to the absence of a linear term in
$N-N_{c}$ in Eqs.~(\ref{eq_dlc-rf},\ref{eq_dlc-ra}). (In the pure
$O(N)$ model on the contrary, $d_{lc}=2- b(2-N) + O((N-2)^{2})$, where
$b=1/4$ is proportional to the temperature of the Kosterlitz-Thouless
transition.\cite{cardy80}) The slope of the curve $N_{lc}(d)$ giving
the locus of the lower critical dimension for QLRO is infinite as
$d\rightarrow 4^{-}$: see Figure~\ref{fig_nc_comp_qlro}.

\begin{figure}[htb]
  \centering
  \includegraphics[width= \linewidth]{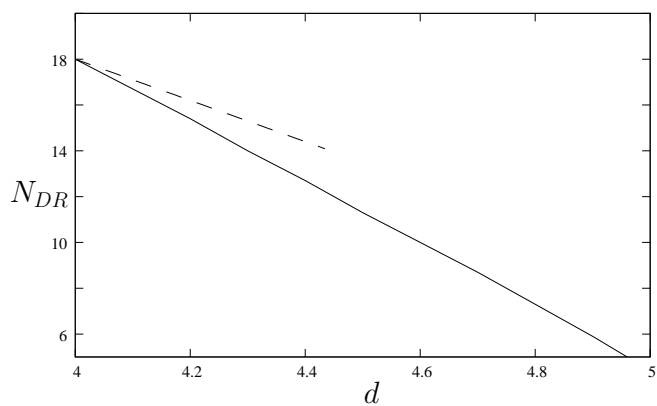}
  \caption{Comparison between the results of the 2-loop perturbative
    FRG (dashed line) and of the NP-FRG (continuous line) near $d=4$
    for $N_{DR}(d)$ in the case of the RF$O(N)$M.  }
  \label{fig_nc_comp_dr}
\end{figure}

The 2-loop predictions around $N=N_{c}$ and $N=N_{lc}(d)$ near $d=4$
are in agreement with those of our NP-FRG treatment. For a detailed
comparison, we plot the two sets of results in
Figures~(\ref{fig_nc_comp_qlro},\ref{fig_nc_comp_dr}). The
nonperturbative, but approximate FRG computation does not exactly
reproduces the 2-loop calculation near $d=4$, but the differences are
not very significant and do not alter the general picture.

To summarize: We have shown in this paper that the theory describing
the long-distance physics of the RF and RA $O(N)$ models near $d=4$ is
perturbatively renormalizable at two loops, thereby proving that the
1-loop result is not fortuitous. The results we have obtained within
the 2-loop order fit into the general scenario predicted by our NP-FRG
approach.\cite{tarjus04,tissier06} Considering the technical
difficulties associated with the FRG loop expansion in $d=4+\epsilon$,
it is highly unlikely that perturbative FRG will ever provide accurate
extrapolations to the physical cases $d=2,3$ (and to $N=1$) for the RF
and RA models. The NP-FRG on the other hand offers a direct way to
study these situations.

\appendix
\section{Fixed points and their stability for the RA$O(N)$M
 in the large $N$ limit}
\label{sec_ra_large_n}

We first rewrite the 1-loop beta function for the derivative of
$\widetilde{T}(\phi)=(N-2)\widetilde{R}(z)$, Eq. (\ref{eq_sp_inf}), by
introducing the function $U(\phi)=-\widetilde{T}'(\phi)/\sin\phi$,
\begin{equation}
  \label{eqA1}
  \begin{split}
  \epsilon^{-1}\sin(\phi) \beta_U(\phi)=&(U(0)-1)\sin(\phi)
  U(\phi)+\\&(U(0)\cos(\phi) -U(\phi)) U'(\phi),
      \end{split}
\end{equation}
and we look for fixed-point solutions. When $U_{\ast}(0)=1$, the only
solutions are $U_{\ast}(\phi)=\cos\phi$ and $U_{\ast}(\phi)=1$. If
$U_{\ast}(0)\neq1$, the equation $\beta_{U}(\phi)=0$ can be solved by
inverting the relation between $U$ and $\phi$ and considering $\phi$
as a function of $U$. The fixed-point equation now reads
\begin{equation}
  \label{eq_flow_ra}
  \partial_U\left (\cos \phi(U)\right
  )-\frac{U_0}{U_0-1}\frac{\cos(\phi(U))} U=-\frac1{U_0-1}
\end{equation}
where $U_{0}$ is such that $\phi(U_{0})=0$. The solutions of
Eq.~(\ref{eq_flow_ra}) are easily found as $\cos(\phi(U))=K
U^{\frac{U_{0}}{U_{0}-1}}+U$. $K$ is a constant that is determined
through the condition $\phi(U_{0})=0$. The result can be reexpressed
by stating that the fixed-point functions $U_{\ast}(\phi)$ are
solutions of the following transcendental equation:
\begin{equation}
  \label{eq_trans}
  U_\ast(\phi)-(U_\ast(0)-1)\left(\frac{U_\ast(\phi)} {U_\ast(0)}
  \right)^{\frac {U_\ast(0)}{U_\ast(0)-1}}=\cos(\phi) 
\end{equation}
where $U_{\ast}(0)$ is different from $1$ but still unknown.

Note that the property $\widetilde{T}(\pi-\phi)=\widetilde{T}(\phi)$ imposes
$U(\pi-\phi)=-U(\phi)$. Since we expect the functions to be analytic
around $\frac{\pi}{2}$ (\ie{}, $z=0$), this property implies
that $U(\frac{\pi}{2})$ and all even derivatives of $U$ evaluated in
$\frac{\pi}{2}$ are equal to zero. This requirement can only be
fulfilled by the solutions of Eq.~(\ref{eq_trans}) if
$\frac{U_{\ast}(0)}{U_{\ast}(0)-1}$ is a nonzero integer. On the other
hand, it cannot be satisfied by the solution $U_{\ast}(\phi)=1$, which
should therefore be discarded.

To determine the acceptable values of $U_{\ast}(0)$, we have to turn
to the stability analysis. Introducing a small deviation around the
fixed point, $\delta(\phi)=U(\phi)-U_{\ast}(\phi)$, and linearizing the
associated FRG equation leads to the following eigenvalue equation:
\begin{equation}
  \label{eq_stab_ra}
  \begin{split}
    (\epsilon^{-1}\Lambda)\sin\phi&\ \delta(\phi)=(\sin\phi\ 
    U_\ast(\phi)+ \cos\phi\ U_\ast'(\phi) )\delta(0)\\&+
    \left((U_\ast(0)-1)\sin\phi-U_\ast'(\phi)\right)\delta(\phi)\\&+(U_\ast(0)\cos\phi-U_\ast(\phi))\delta'(\phi) .
  \end{split}
\end{equation}

We first check that the solution $U_{\ast}(\phi)=\cos\phi$,
corresponding to $U_{\ast}(0)=1$, is fully unstable with
$\epsilon^{-1}\Lambda=1$, and we next consider the solutions given by
Eq.~(\ref{eq_trans}). As before we consider $\phi$ as a function of
$U\equiv U_{\ast}(\phi)$, which gives after some manipulations,
\begin{equation}
  \label{eq_ra_sol_i}
  \begin{split}
    &\left(\frac U{U_0}\right) ^{\frac{U_0}{U_0-1}}\delta(U_0)= -
    (U_0-1)\left(1-\left(\frac U{U_0}\right)
      ^{\frac 1{U_0-1}}\right)U\delta'(U)\\&+
    \left[1-(\epsilon^{-1}\Lambda-U_0+1)\left(1-\left(\frac U{U_0}\right)
        ^{\frac 1{U_0-1}}\right)\right]\delta(U)
  \end{split}
\end{equation}
where $U_{0}\equiv U_{\ast}(\phi=0)$. The solution of the above
equations is easily obtained as:
\begin{equation}
  \label{eq_sol_ev_ra}
  \begin{split}
    \delta(U)=\frac{\delta(U_0)}{\epsilon^{-1}\Lambda}&\left(\frac
      U{U_0}\right) ^{\frac
      {U_0-\epsilon^{-1}\Lambda}{U_0-1}}\left(1-\left(\frac
        U{U_0}\right) ^{\frac
        {\epsilon^{-1}\Lambda}{U_0-1}}\right)\\&\left(1-\left(\frac
        U{U_0}\right) ^{\frac {1}{U_0-1}}\right)^{-1}
  \end{split}
\end{equation}
where, we recall, $U_{0}/(U_{0}-1)$ is a nonzero integer. The
condition that $\delta(U)$ be an odd fuction of $U$, analytic around
$U=0$, imposes stringent constraints on $\epsilon^{-1}\Lambda$. One
finds that the only singly unstable fixed point corresponds to
$U_{0}/(U_{0}-1)=3$, \ie{}, $U_{\ast}(0)=3/2$. The associated
eigenvalues are equal to $\epsilon^{-1}\Lambda=1,0,-1,-2,-3,...$. For
$U_{\ast}(0)=3/2$, Eq.~(\ref{eq_trans}) can be explicitly solved,
which leads to $U_{\ast}(\phi)=3\cos(\frac{\pi+\phi}{3})$ and to
Eq.~(\ref{eq_Sp_fp}).

To derive the fixed-point solution at the following orders in $1/N$, great simplification is obtained by first noticing that the above
solution (in the limit $N\rightarrow \infty$) can be rewritten as
\begin{equation}
  \label{eq_sol_ra_2}
  \widetilde
  T_{\ast}'(\phi)=-\dfrac{3}{2} \sin(\dfrac{\pi-2\phi}{3}) \left[2\cos(\dfrac{\pi-2\phi}{3})-1\right],
\end{equation}
which implies that $\widetilde T_{\ast}(\phi)$ is a function of
$\cos(\frac{\pi-2\phi}{3})$. For finite $N$ we now look for a
fixed-point solution of the form $\widetilde T_{\ast}'(\phi)=-\frac{3}{2}
\sin(\frac{\pi-2\phi}{3})(2X-1)G(X)$ with
$X=\cos(\frac{\pi-2\phi}{3})$. $G(X)$ can be expanded in powers of
$1/N$ (or for convenience, $1/(N-2)$), its leading term being simply
equal to 1 (see Eq.~(\ref{eq_sol_ra_2})). With this transformation the
fixed-point equation can be solved in powers of $1/(N-2)$, each term being a
polynomial in $X$. One obtains for the first terms
\begin{equation}
  \label{eqA9}
  \begin{split}
  G(X)=1+ \dfrac{2}{9(N-2)}&\left(95-44X-16X^{2}\right)\\&+ 
\mathcal O \left(\dfrac{1}{(N-2)^{2}}\right).
      \end{split}
\end{equation}

From this expression, one derives
$-T_{\ast}''(0)=\frac{3}{2}G(\frac{1}{2})=\frac{3}{2}+
\frac{23}{N}+O(\frac{1}{N^{2}})$, which leads to the expressions given
in section~\ref{sec_ra}.

Finally, the fixed point can also be found at the 2-loop level in the
large $N$ limit by using the above variable $X$. One obtains for
instance that the correlation length exponent is equal to
\begin{equation}
  \label{eq_nu_ra_large_n}
  \nu=\frac 1\epsilon-\frac{17}{3N}-\frac{29707}{81 N^2}+\mathcal
  O\left(\frac 1{N^3},\epsilon \right),
\end{equation}
so that it is now different from the dimensional-reduction value.

\section{2-loops diagrams}
\label{sec_diagrams}
We give here all the 2-loop diagrams built with the rules defined in section IV A.

\begin{figure*}[htbp]
  \raggedright \includegraphics[width=.094 \linewidth, bb=145 605 265
  690]{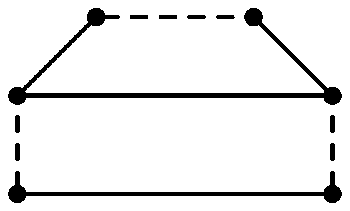} \includegraphics[width=.094 \linewidth, bb=145 605
  265 690]{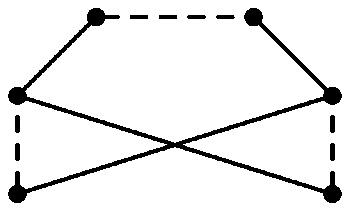} \includegraphics[width=.094 \linewidth, bb=145
  605 265 690]{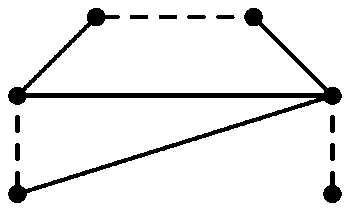} \includegraphics[width=.094 \linewidth,
  bb=145 605 265 690]{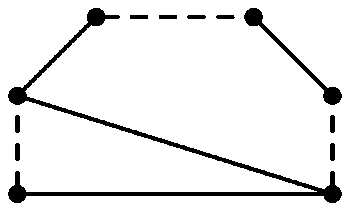} \includegraphics[width=.094
  \linewidth, bb=145 605 265 690]{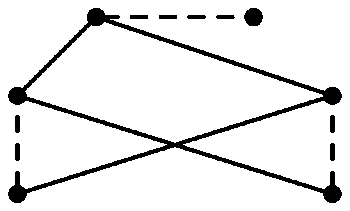}
  \includegraphics[width=.094 \linewidth, bb=145 605 265
  690]{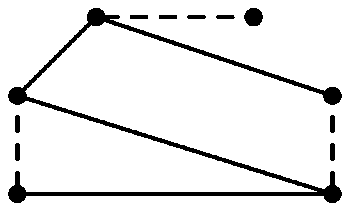} \includegraphics[width=.094 \linewidth, bb=145 605
  265 690]{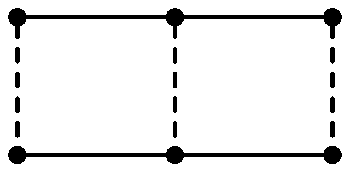} \includegraphics[width=.094 \linewidth, bb=145
  605 265 690]{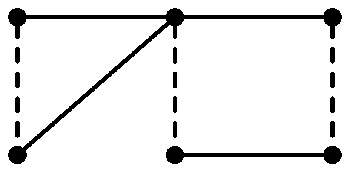} \includegraphics[width=.094 \linewidth,
  bb=145 605 265 690]{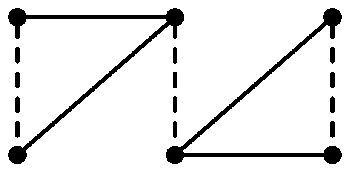} \includegraphics[width=.094
  \linewidth, bb=145 605 265 690]{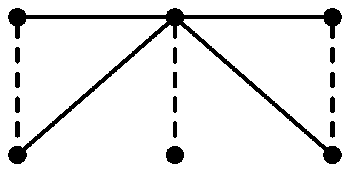}
  \includegraphics[width=.094 \linewidth, bb=145 605 265
  690]{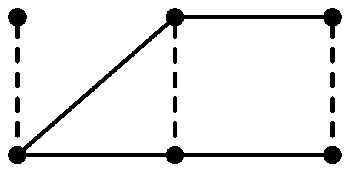} \includegraphics[width=.094 \linewidth, bb=145 605
  265 690]{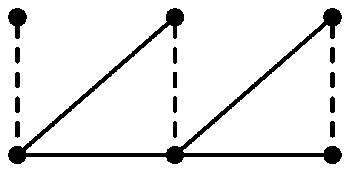} \includegraphics[width=.094 \linewidth, bb=145
  605 265 690]{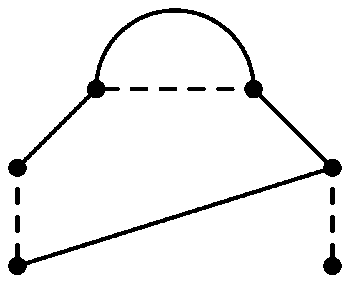} \includegraphics[width=.094 \linewidth,
  bb=145 605 265 690]{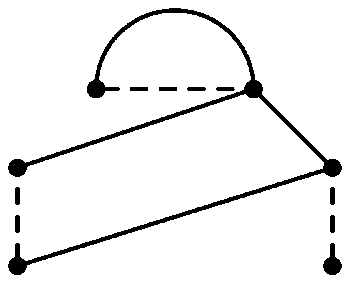} \includegraphics[width=.094
  \linewidth, bb=145 605 265 690]{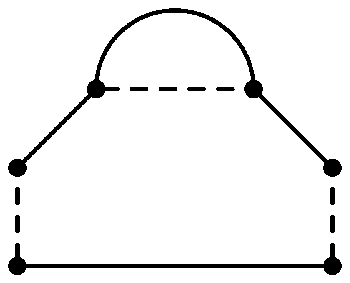}
  \includegraphics[width=.094 \linewidth, bb=145 605 265
  690]{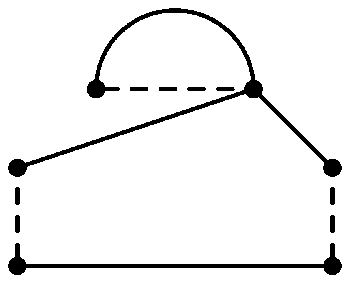} \includegraphics[width=.094 \linewidth, bb=145 605
  265 690]{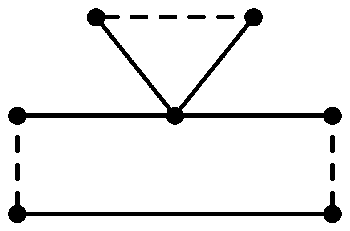} \includegraphics[width=.094 \linewidth, bb=145
  605 265 690]{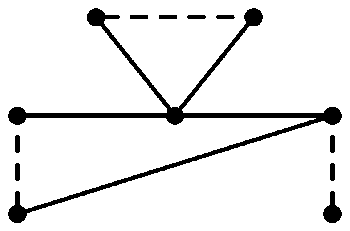} \includegraphics[width=.094 \linewidth,
  bb=145 605 265 690]{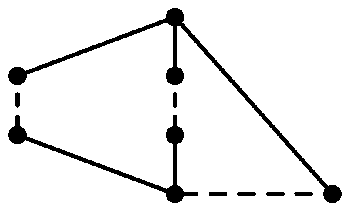} \includegraphics[width=.094
  \linewidth, bb=145 605 265 690]{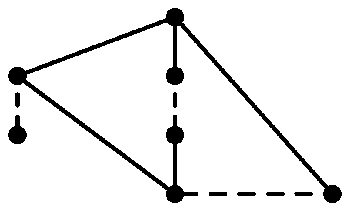}
  \includegraphics[width=.094 \linewidth, bb=145 605 265
  690]{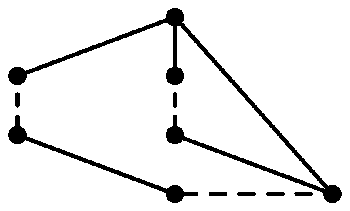} \includegraphics[width=.094 \linewidth, bb=145 605
  265 690]{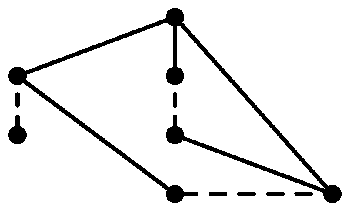} \includegraphics[width=.094 \linewidth, bb=145
  605 265 690]{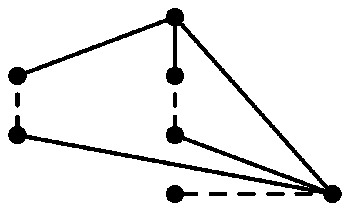} \includegraphics[width=.094 \linewidth,
  bb=145 605 265 690]{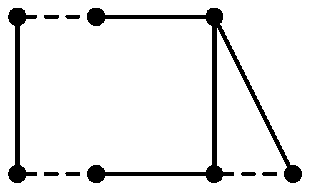} \includegraphics[width=.094
  \linewidth, bb=145 605 265 690]{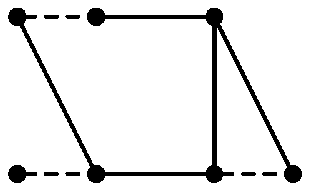}
  \includegraphics[width=.094 \linewidth, bb=145 605 265
  690]{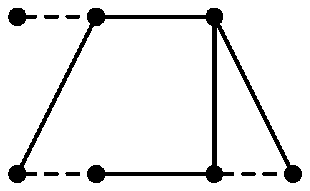} \includegraphics[width=.094 \linewidth, bb=145 605
  265 690]{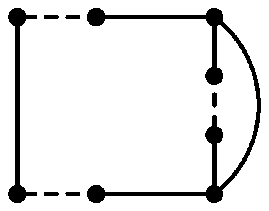} \includegraphics[width=.094 \linewidth, bb=145
  605 265 690]{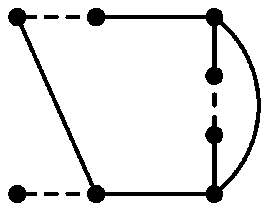} \includegraphics[width=.094 \linewidth,
  bb=145 605 265 690]{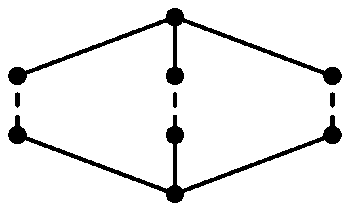} \includegraphics[width=.094
  \linewidth, bb=145 605 265 690]{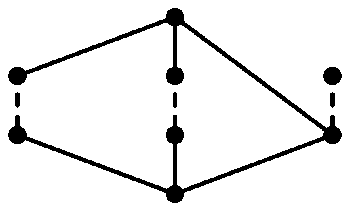}
\caption{2-loop diagrams for the  2-replica effective action.} 
\label{fig_diag_2loops}
\end{figure*}
\begin{figure*}[htbp]
\raggedright
  \includegraphics[width=.094 \linewidth, bb=145  605  265  670]{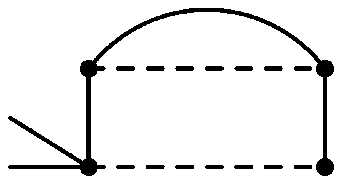}
  \includegraphics[width=.094 \linewidth, bb=145  605  265  670]{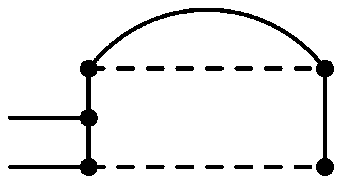}
  \includegraphics[width=.094 \linewidth, bb=145  605  265  670]{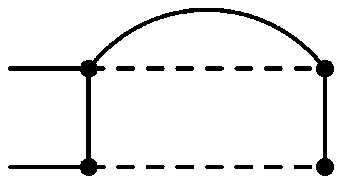}
  \includegraphics[width=.094 \linewidth, bb=145  605  265  670]{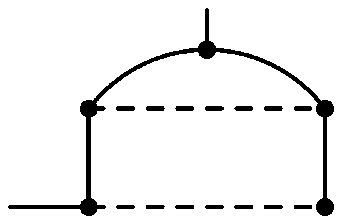}
  \includegraphics[width=.094 \linewidth, bb=145  605  265  670]{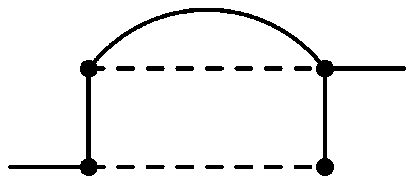}
  \includegraphics[width=.094 \linewidth, bb=145  605  265  670]{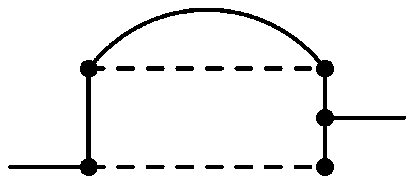}
  \includegraphics[width=.094 \linewidth, bb=145  605  265  670]{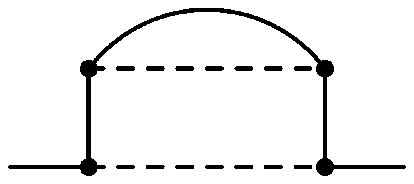}
  \includegraphics[width=.094 \linewidth, bb=145  605  265  670]{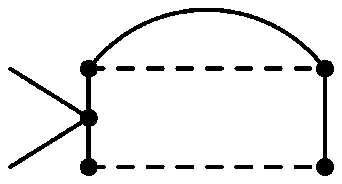}
  \includegraphics[width=.094 \linewidth, bb=145  605  265  670]{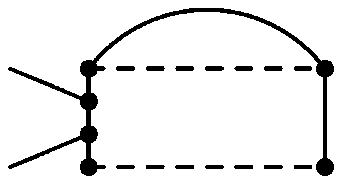}
  \includegraphics[width=.094 \linewidth, bb=145  605  265  670]{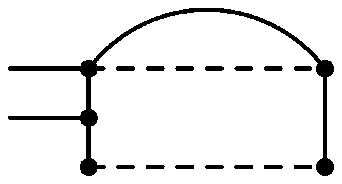}
  \includegraphics[width=.094 \linewidth, bb=145  605  265  670]{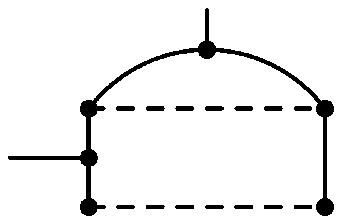}
  \includegraphics[width=.094 \linewidth, bb=145  605  265  670]{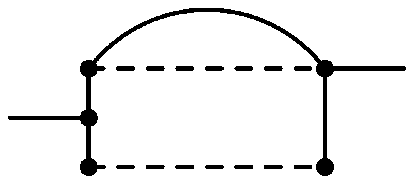}
  \includegraphics[width=.094 \linewidth, bb=145  605  265  670]{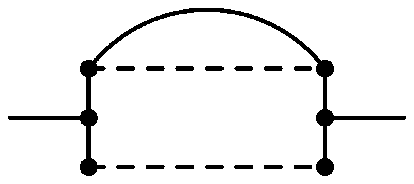}
  \includegraphics[width=.094 \linewidth, bb=145  605  265  670]{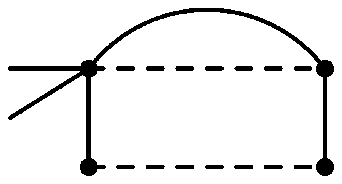}
  \includegraphics[width=.094 \linewidth, bb=145  605  265  670]{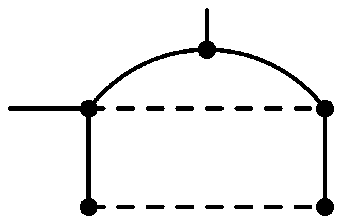}
  \includegraphics[width=.094 \linewidth, bb=145  605  265  670]{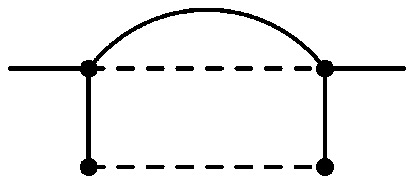}
  \includegraphics[width=.094 \linewidth, bb=145  605  265  670]{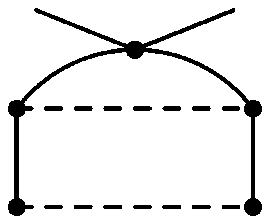}
  \includegraphics[width=.094 \linewidth, bb=145  605  265  670]{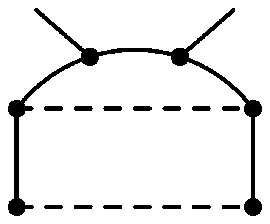}
\caption{2-loop diagrams of type A for the 1-replica 2-point proper vertex.} 
\label{fig_diag_B}
\end{figure*}
\begin{figure*}[htbp]
\raggedright
  \includegraphics[width=.094 \linewidth, bb=145  605  265  670]{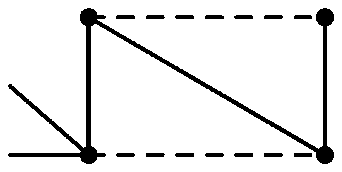}
  \includegraphics[width=.094 \linewidth, bb=145  605  265  670]{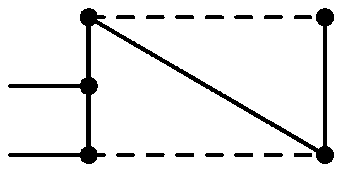}
  \includegraphics[width=.094 \linewidth, bb=145  605  265  670]{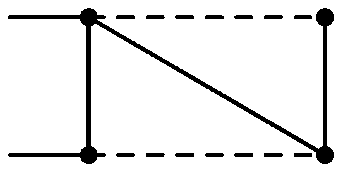}
  \includegraphics[width=.094 \linewidth, bb=145  605  265  670]{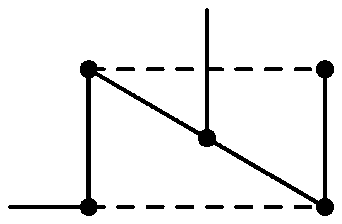}
  \includegraphics[width=.094 \linewidth, bb=145  605  265  670]{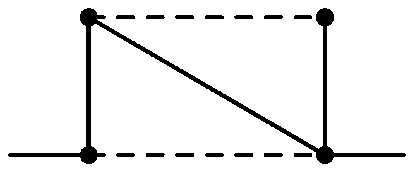}
  \includegraphics[width=.094 \linewidth, bb=145  605  265  670]{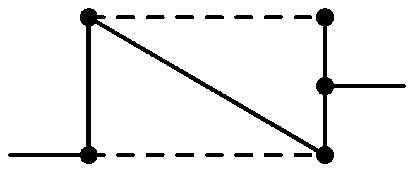}
  \includegraphics[width=.094 \linewidth, bb=145  605  265  670]{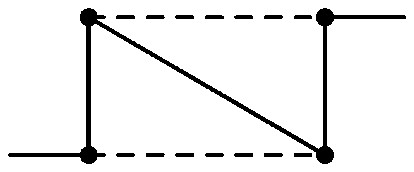}
  \includegraphics[width=.094 \linewidth, bb=145  605  265  670]{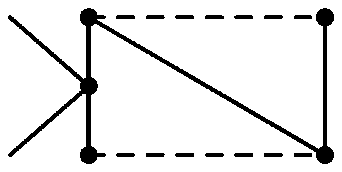}
  \includegraphics[width=.094 \linewidth, bb=145  605  265  670]{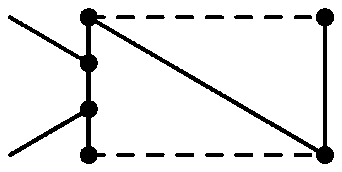}
  \includegraphics[width=.094 \linewidth, bb=145  605  265  670]{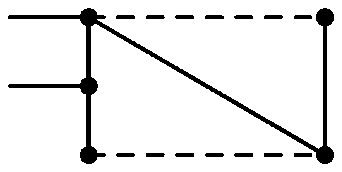}
  \includegraphics[width=.094 \linewidth, bb=145  605  265  670]{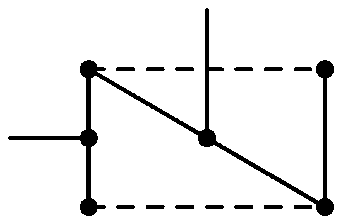}
  \includegraphics[width=.094 \linewidth, bb=145  605  265  670]{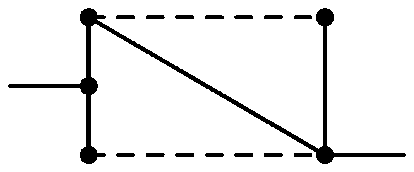}
  \includegraphics[width=.094 \linewidth, bb=145  605  265  670]{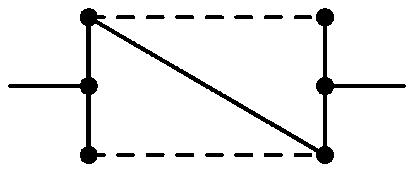}
  \includegraphics[width=.094 \linewidth, bb=145  605  265  670]{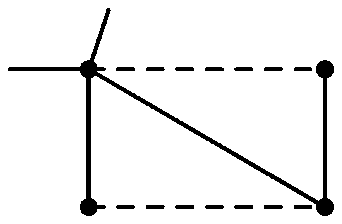}
  \includegraphics[width=.094 \linewidth, bb=145  605  265  670]{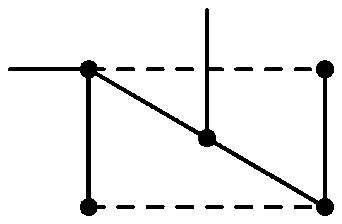}
  \includegraphics[width=.094 \linewidth, bb=145  605  265  670]{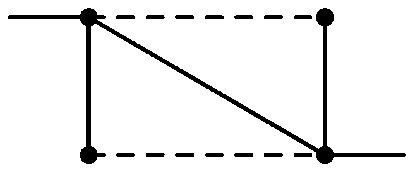}
  \includegraphics[width=.094 \linewidth, bb=145  605  265  670]{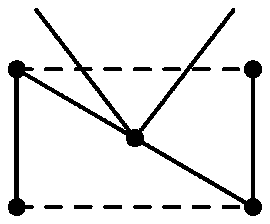}
  \includegraphics[width=.094 \linewidth, bb=145  605  265  670]{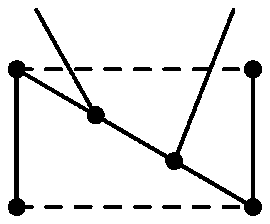}
\caption{2-loop diagrams of type B for the 1-replica 2-point proper vertex.} 
\label{fig_diag_C}
\end{figure*}
\begin{figure*}[htbp]
\raggedright
  \includegraphics[width=.094 \linewidth, bb=145  605  265  670]{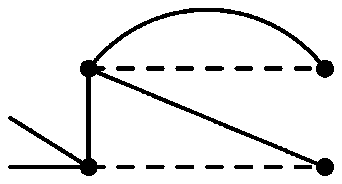}
  \includegraphics[width=.094 \linewidth, bb=145  605  265  670]{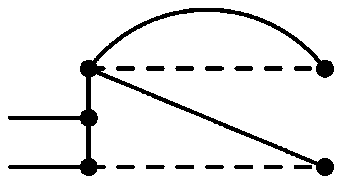}
  \includegraphics[width=.094 \linewidth, bb=145  605  265  670]{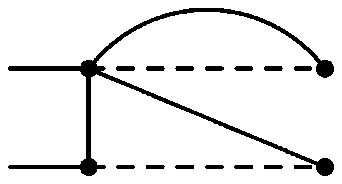}
  \includegraphics[width=.094 \linewidth, bb=145  605  265  670]{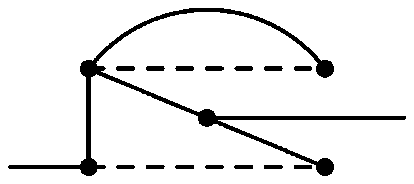}
  \includegraphics[width=.094 \linewidth, bb=145  605  265  670]{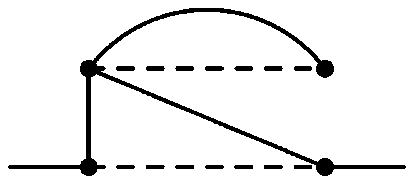}
  \includegraphics[width=.094 \linewidth, bb=145  605  265  670]{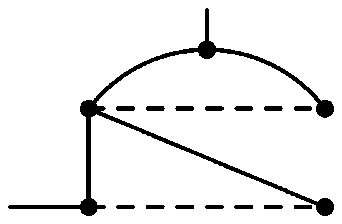}
  \includegraphics[width=.094 \linewidth, bb=145  605  265  670]{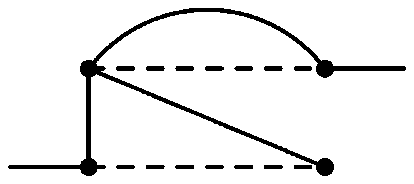}
  \includegraphics[width=.094 \linewidth, bb=145  605  265  670]{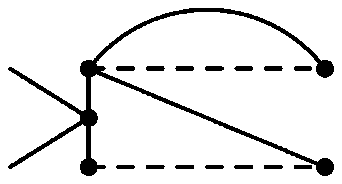}
  \includegraphics[width=.094 \linewidth, bb=145  605  265  670]{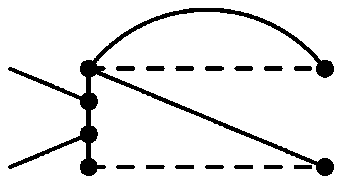}
  \includegraphics[width=.094 \linewidth, bb=145  605  265  670]{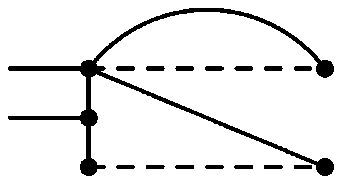}
  \includegraphics[width=.094 \linewidth, bb=145  605  265  670]{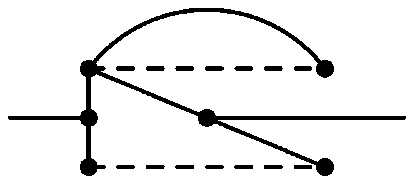}
  \includegraphics[width=.094 \linewidth, bb=145  605  265  670]{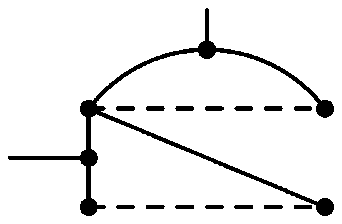}
  \includegraphics[width=.094 \linewidth, bb=145  605  265  670]{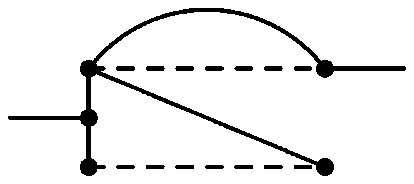}
  \includegraphics[width=.094 \linewidth, bb=145  605  265  670]{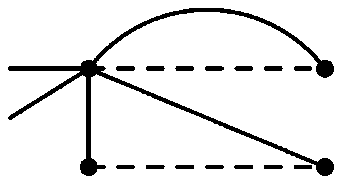}
  \includegraphics[width=.094 \linewidth, bb=145  605  265  670]{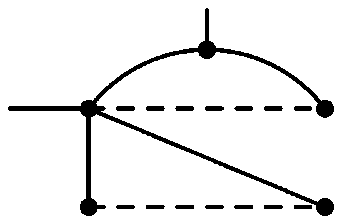}
  \includegraphics[width=.094 \linewidth, bb=145  605  265  670]{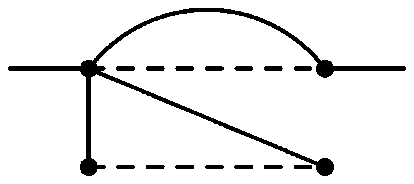}
  \includegraphics[width=.094 \linewidth, bb=145  605  265  670]{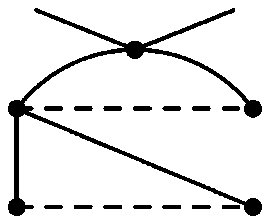}
  \includegraphics[width=.094 \linewidth, bb=145  605  265  670]{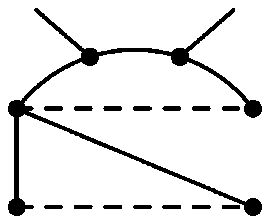}
  \includegraphics[width=.094 \linewidth, bb=145  605  265  670]{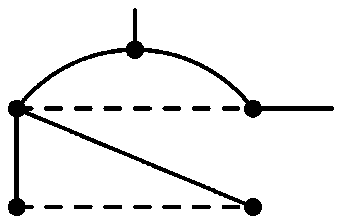}
  \includegraphics[width=.094 \linewidth, bb=145  605  265  670]{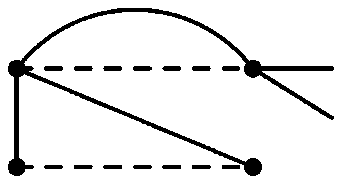}
\caption{2-loop diagrams of type C for the 1-replica 2-point proper vertex.} 
\label{fig_diag_D}
\end{figure*}
\begin{figure*}[htbp]
\raggedright
  \includegraphics[width=.094 \linewidth, bb=145  605  265  670]{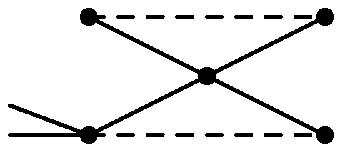}
  \includegraphics[width=.094 \linewidth, bb=145  605  265  670]{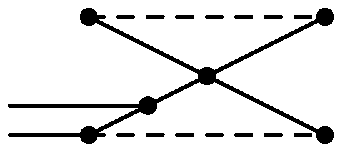}
  \includegraphics[width=.094 \linewidth, bb=145  605  265  670]{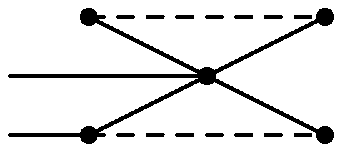}
  \includegraphics[width=.094 \linewidth, bb=145  605  265  670]{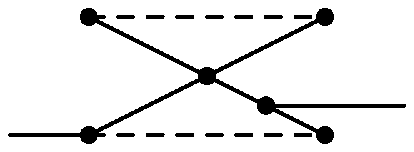}
  \includegraphics[width=.094 \linewidth, bb=145  605  265  670]{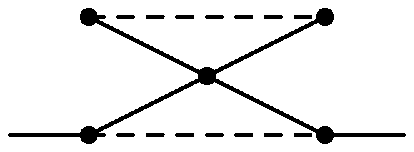}
  \includegraphics[width=.094 \linewidth, bb=145  605  265  670]{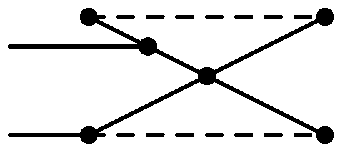}
  \includegraphics[width=.094 \linewidth, bb=145  605  265  670]{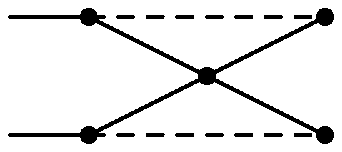}
  \includegraphics[width=.094 \linewidth, bb=145  605  265  670]{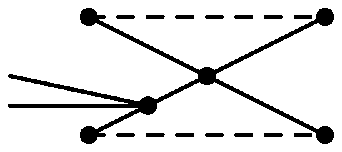}
  \includegraphics[width=.094 \linewidth, bb=145  605  265  670]{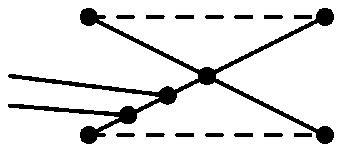}
  \includegraphics[width=.094 \linewidth, bb=145  605  265  670]{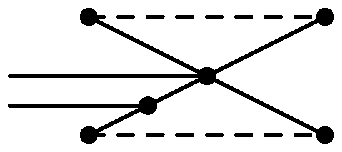}
  \includegraphics[width=.094 \linewidth, bb=145  605  265  670]{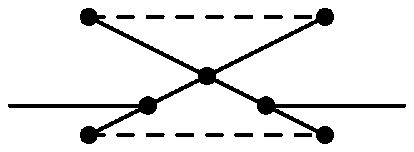}
  \includegraphics[width=.094 \linewidth, bb=145  605  265  670]{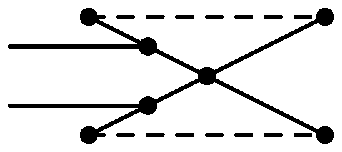}
  \includegraphics[width=.094 \linewidth, bb=145  605  265  670]{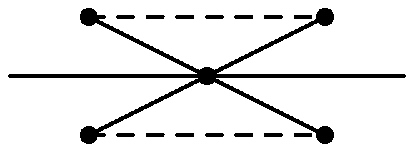}
\caption{2-loop diagrams of type D for the 1-replica 2-point proper vertex.} 
\label{fig_diag_E}
\end{figure*}
\begin{figure*}[htbp]
\raggedright
  \includegraphics[width=.094 \linewidth, bb=145  605  265  670]{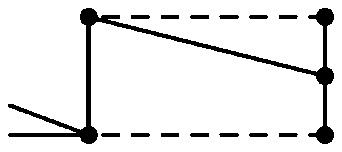}
  \includegraphics[width=.094 \linewidth, bb=145  605  265  670]{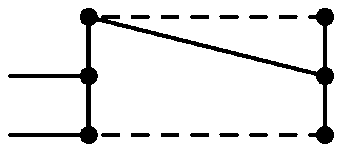}
  \includegraphics[width=.094 \linewidth, bb=145  605  265  670]{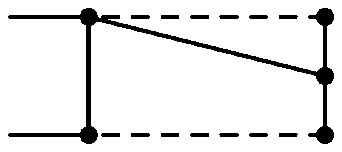}
  \includegraphics[width=.094 \linewidth, bb=145  605  265  670]{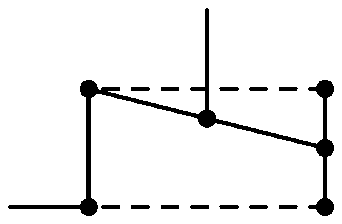}
  \includegraphics[width=.094 \linewidth, bb=145  605  265  670]{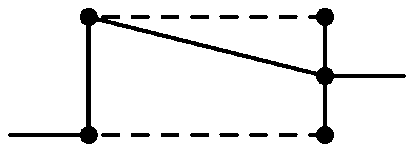}
  \includegraphics[width=.094 \linewidth, bb=145  605  265  670]{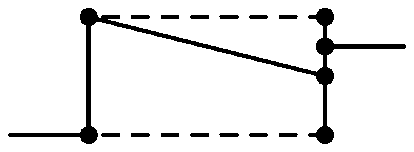}
  \includegraphics[width=.094 \linewidth, bb=145  605  265  670]{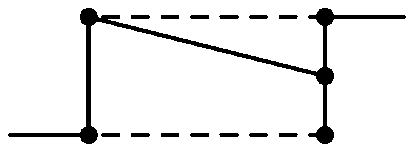}
  \includegraphics[width=.094 \linewidth, bb=145  605  265  670]{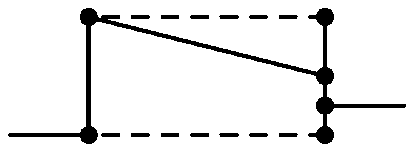}
  \includegraphics[width=.094 \linewidth, bb=145  605  265  670]{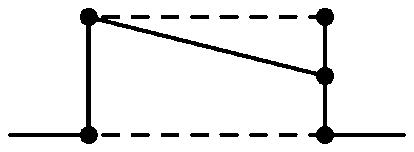}
  \includegraphics[width=.094 \linewidth, bb=145  605  265  670]{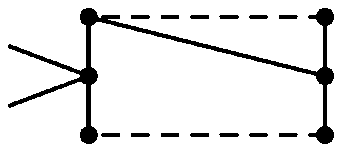}
  \includegraphics[width=.094 \linewidth, bb=145  605  265  670]{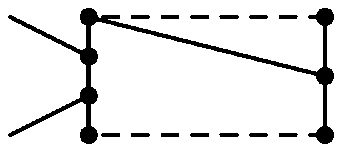}
  \includegraphics[width=.094 \linewidth, bb=145  605  265  670]{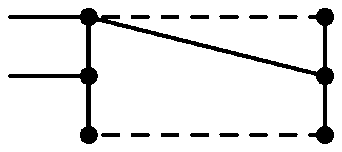}
  \includegraphics[width=.094 \linewidth, bb=145  605  265  670]{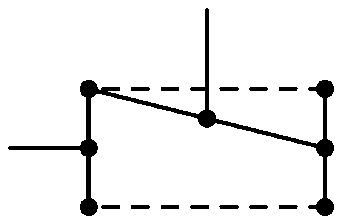}
  \includegraphics[width=.094 \linewidth, bb=145  605  265  670]{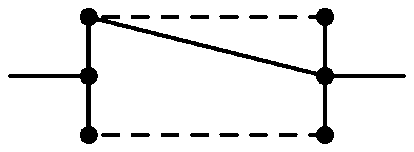}
  \includegraphics[width=.094 \linewidth, bb=145  605  265  670]{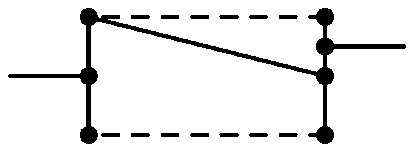}
  \includegraphics[width=.094 \linewidth, bb=145  605  265  670]{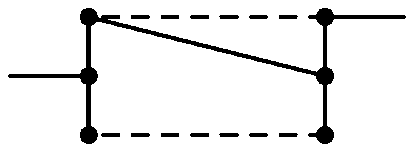}
  \includegraphics[width=.094 \linewidth, bb=145  605  265  670]{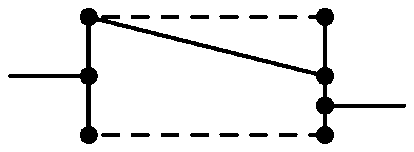}
  \includegraphics[width=.094 \linewidth, bb=145  605  265  670]{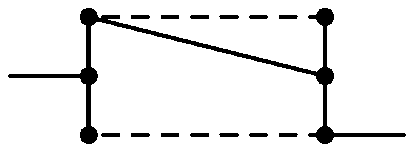}
  \includegraphics[width=.094 \linewidth, bb=145  605  265  670]{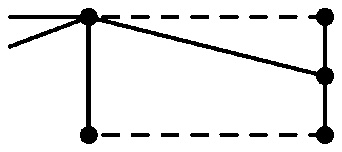}
  \includegraphics[width=.094 \linewidth, bb=145  605  265  670]{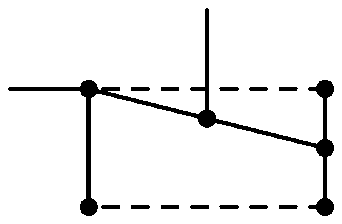}
  \includegraphics[width=.094 \linewidth, bb=145  605  265  670]{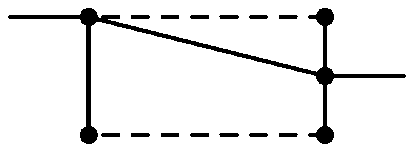}
  \includegraphics[width=.094 \linewidth, bb=145  605  265  670]{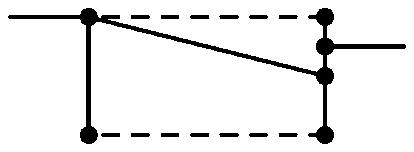}
  \includegraphics[width=.094 \linewidth, bb=145  605  265  670]{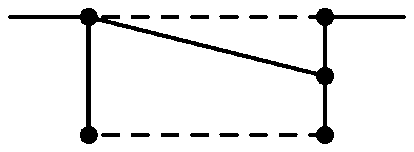}
  \includegraphics[width=.094 \linewidth, bb=145  605  265  670]{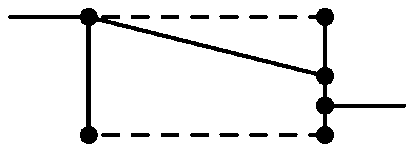}
  \includegraphics[width=.094 \linewidth, bb=145  605  265  670]{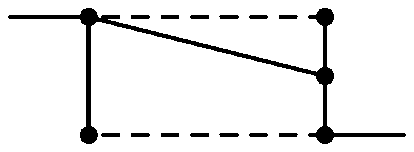}
  \includegraphics[width=.094 \linewidth, bb=145  605  265  670]{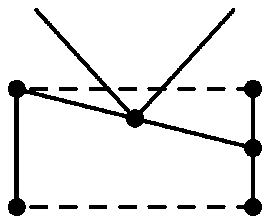}
  \includegraphics[width=.094 \linewidth, bb=145  605  265  670]{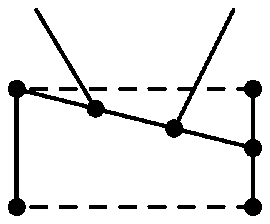}
  \includegraphics[width=.094 \linewidth, bb=145  605  265  670]{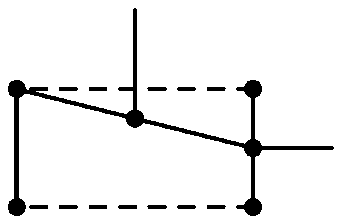}
  \includegraphics[width=.094 \linewidth, bb=145  605  265  670]{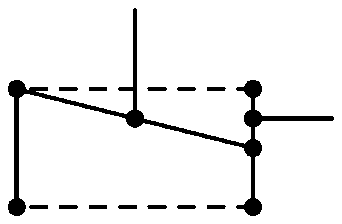}
  \includegraphics[width=.094 \linewidth, bb=145  605  265  670]{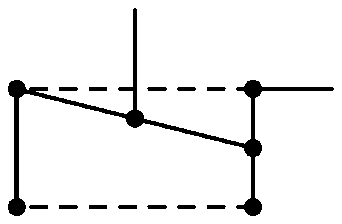}
  \includegraphics[width=.094 \linewidth, bb=145  605  265  670]{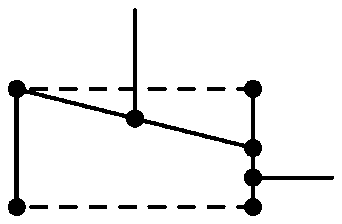}
  \includegraphics[width=.094 \linewidth, bb=145  605  265  670]{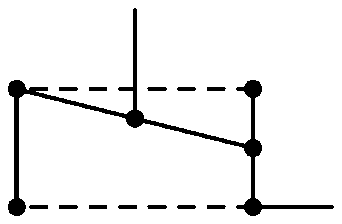}
  \includegraphics[width=.094 \linewidth, bb=145  605  265  670]{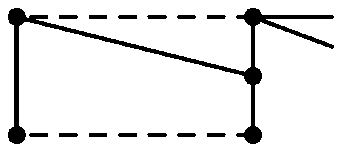}
  \includegraphics[width=.094 \linewidth, bb=145  605  265  670]{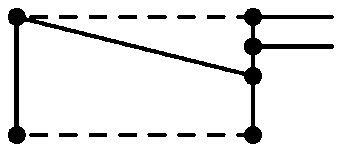}
  \includegraphics[width=.094 \linewidth, bb=145  605  265  670]{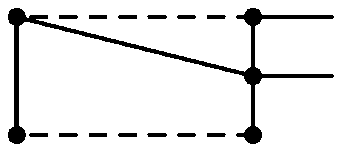}
  \includegraphics[width=.094 \linewidth, bb=145  605  265  670]{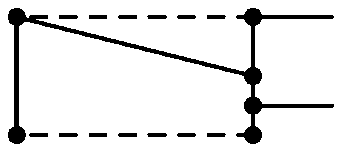}
  \includegraphics[width=.094 \linewidth, bb=145  605  265  670]{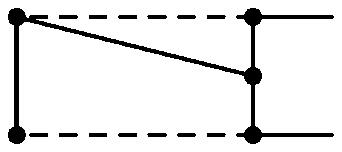}
  \includegraphics[width=.094 \linewidth, bb=145  605  265  670]{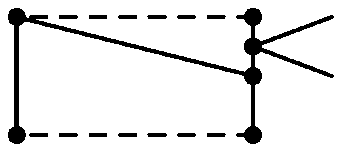}
  \includegraphics[width=.094 \linewidth, bb=145  605  265  670]{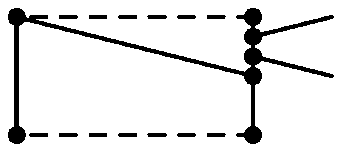}
  \includegraphics[width=.094 \linewidth, bb=145  605  265  670]{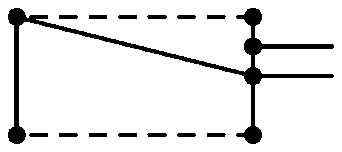}
  \includegraphics[width=.094 \linewidth, bb=145  605  265  670]{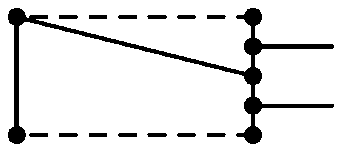}
  \includegraphics[width=.094 \linewidth, bb=145  605  265  670]{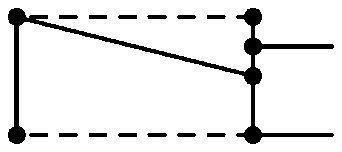}
  \includegraphics[width=.094 \linewidth, bb=145  605  265  670]{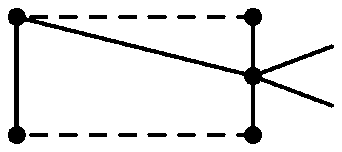}
  \includegraphics[width=.094 \linewidth, bb=145  605  265  670]{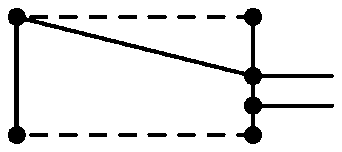}
  \includegraphics[width=.094 \linewidth, bb=145  605  265  670]{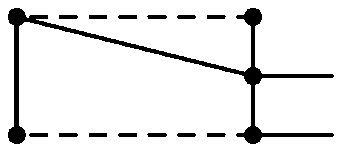}
  \includegraphics[width=.094 \linewidth, bb=145  605  265  670]{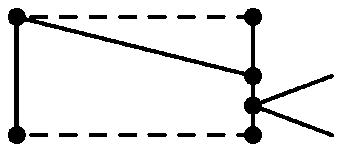}
  \includegraphics[width=.094 \linewidth, bb=145  605  265  670]{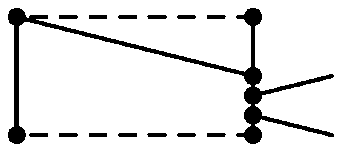}
  \includegraphics[width=.094 \linewidth, bb=145  605  265  670]{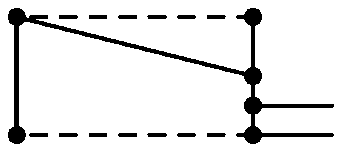}
  \includegraphics[width=.094 \linewidth, bb=145  605  265  670]{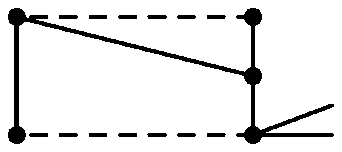}
\caption{2-loop diagrams of type E for the 1-replica 2-point proper vertex.} 
\label{fig_diag_F}
\end{figure*}
\begin{figure*}[htbp]
 \raggedright
 \includegraphics[width=.094 \linewidth, bb=145  605  265  670]{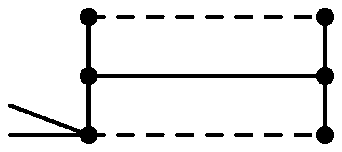}
  \includegraphics[width=.094 \linewidth, bb=145  605  265  670]{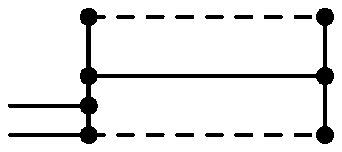}
  \includegraphics[width=.094 \linewidth, bb=145  605  265  670]{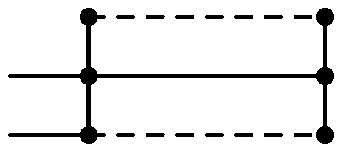}
  \includegraphics[width=.094 \linewidth, bb=145  605  265  670]{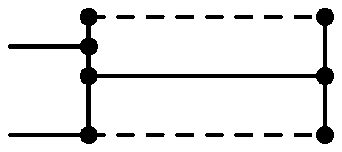}
  \includegraphics[width=.094 \linewidth, bb=145  605  265  670]{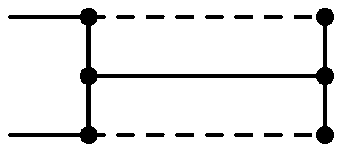}
  \includegraphics[width=.094 \linewidth, bb=145  605  265  670]{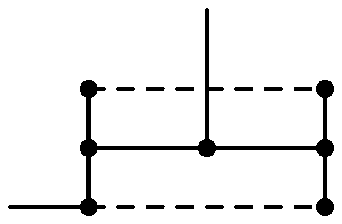}
  \includegraphics[width=.094 \linewidth, bb=145  605  265  670]{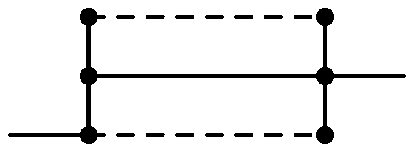}
  \includegraphics[width=.094 \linewidth, bb=145  605  265  670]{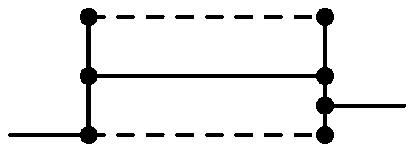}
  \includegraphics[width=.094 \linewidth, bb=145  605  265  670]{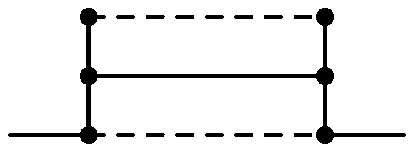}
  \includegraphics[width=.094 \linewidth, bb=145  605  265  670]{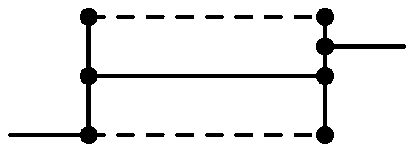}
  \includegraphics[width=.094 \linewidth, bb=145  605  265  670]{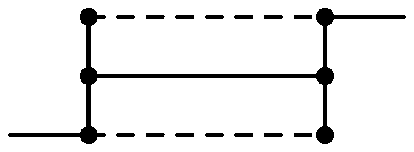}
  \includegraphics[width=.094 \linewidth, bb=145  605  265  670]{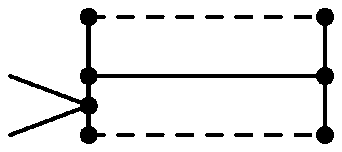}
  \includegraphics[width=.094 \linewidth, bb=145  605  265  670]{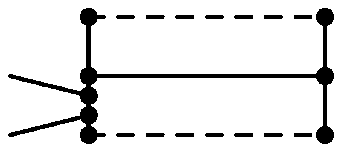}
  \includegraphics[width=.094 \linewidth, bb=145  605  265  670]{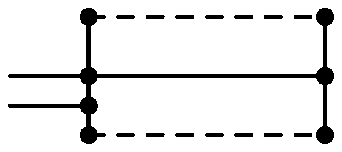}
  \includegraphics[width=.094 \linewidth, bb=145  605  265  670]{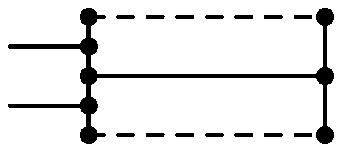}
  \includegraphics[width=.094 \linewidth, bb=145  605  265  670]{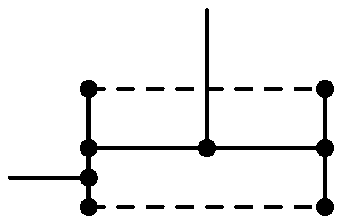}
  \includegraphics[width=.094 \linewidth, bb=145  605  265  670]{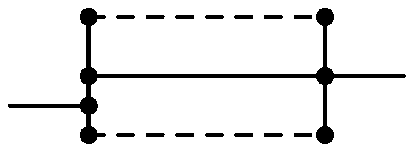}
  \includegraphics[width=.094 \linewidth, bb=145  605  265  670]{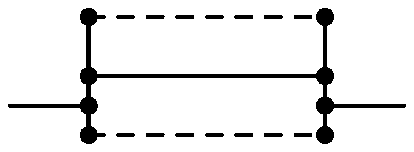}
  \includegraphics[width=.094 \linewidth, bb=145  605  265  670]{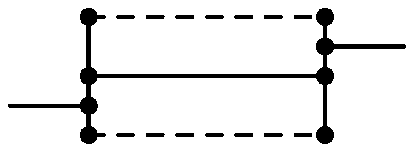}
  \includegraphics[width=.094 \linewidth, bb=145  605  265  670]{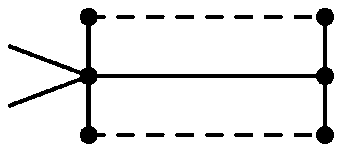}
  \includegraphics[width=.094 \linewidth, bb=145  605  265  670]{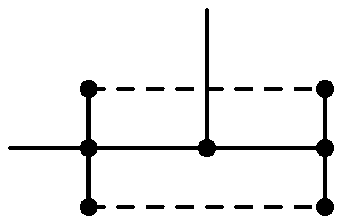}
  \includegraphics[width=.094 \linewidth, bb=145  605  265  670]{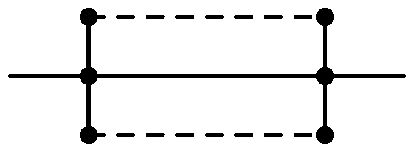}
  \includegraphics[width=.094 \linewidth, bb=145  605  265  670]{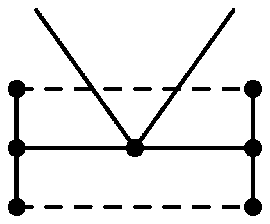}
  \includegraphics[width=.094 \linewidth, bb=145  605  265  670]{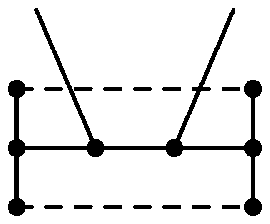}
\caption{2-loop diagrams of type F for the 1-replica 2-point proper vertex.} 
\label{fig_diag_G}
\end{figure*}

\section{2-loop integrals in dimensional regularization}
\label{sec_integrals}

In this section, we discuss the procedure used to evaluate the
integrals appearing in the two-loop calculation in $d=4+\epsilon$.

\subsection{Integrals for the 2-replica diagrams}
 We start with the integrals for the
2-replica effective action. The most general integral reads
\begin{equation}
  \label{eq_int}
  \begin{split}
    &J^{r,s,t}_{i,j,k,l,m,n}(a,b)=\int_{\vect q_1,\vect q_2,\vect
      q_3}\delta(\vect q_1+\vect q_2+\vect q_3)\\& \frac{(\vect
      q_1\cdot \vect q_2)^r(\vect q_2\cdot \vect q_3)^s(\vect q_3\cdot
      \vect q_1)^t}
    {(q_1^2+a)^i(q_2^2+a)^j(q_3^2+a)^k(q_1^2+b)^l(q_2^2+b)^m(q_3^2+b)^n}
      \end{split}
\end{equation}
with the parameters $\{i,j,k,l,m,n,r,s,t\}$ being nonnegative
integers, $\{a,b\}$ being positive real numbers and $r+s+t\leq 2$. The
integrals have obvious symmetry properties since the integration
variables can be exchanged. For instance, the integral is unchanged
when $\{i,l,s\}$ and $\{j,m,t\}$ are exchanged, when $\{j,m,r\}$ and
$\{k,n,t\}$ are exchanged, or when $\{i,j,k,a\}$ and $\{l,m,n,b\}$ are
exchanged.

It is possible to reduce the range of values of $\{r,s,t\}$ that one must consider. This can be done
by rewriting in the integrand $\vect q_1\cdot \vect q_2=
(q_3^2-q_1^2-q_2^2)/2$.  $q_1^2$ can then be replaced by
$(q_1^2+a)-a$, which, if $i>0$, can be combined with
$(q_1^2+a)^{-i}$ to give $(q_1^2+a)^{-i+1}-a(q_1^2+a)^{-i}$. Similar
transformations can be done for $q_2^2$ and $q_3^2$. Then, under the
assumption that $i$, $j$ and $k$ are nonzero, one gets the identity
\begin{equation}
  \begin{split}
  J^{rst}_{i,j,k,l,m,n}=\frac 12(&J^{r-1,s,t}_{i,j,k-1,l,m,n}-
  J^{r-1,s,t}_{i,j-1,k,l,m,n}-\\ &J^{r-1,s,t}_{i-1,j,k,l,m,n}+ a 
J^{r-1,s,t}_{i,j,k,l,m,n}).
     \end{split}
\end{equation}
Similar relations can be obtained under the (weaker) condition that
$i+l$, $j+m$ and $k+n$ are nonzero.

There remains to treat the case where one of the three previous
combinations (say for instance the last one) vanishes (which
implies $k=n=0$). It is sufficient to treat the three cases where
$\{r,s,t\}=\{1,0,0\}$ up to a permutation, because the other
possibilities never appear in the calculation. Consider for instance
$J^{100}_{i,j,0,l,m,0}$. Since $q_3$ appears in the integrand only
through the $\delta$ function, one performs the integral on $\vect
q_3$ trivially. The integrand can then be factorized in a piece $\vect
q_1(q_1^2+a)^{-i}(q_1^2+b)^{-l}$ depending only on $\vect q_1$ and
another (of a similar form) depending only on $\vect q_2$. Each piece
is a vector so that the integral vanishes for symmetry reasons.
Consider now the integral $J^{010}_{i,j,0,l,m,0}$. By integrating
over $\vect q_3$, the numerator of the integrand becomes $-(q_2^2+\vect
q_1\cdot \vect q_2)$. The last term gives zero after integration for
symmetry reasons, just as before. The first term can be rewritten as
$(q_2^2+a)-a$ so that, under the condition that $j>0$
\begin{equation}
  \begin{split}
    J^{010}_{i,j,0,l,m,0}=a J^{000}_{i,j,0,l,m,0}-
    J^{000}_{i,j-1,0,l,m,0}.
     \end{split}
\end{equation}
Similar equations can be obtained if $j=0$ and $m>0$, or if $r=s=0$,
$t=1$.

We can further simplify the integrals by using the relation
\begin{equation}
  \frac 1{(q^2+a)(q^2+b)}=\frac1{a-b}\left(\frac 1{q^2+b} 
-\frac 1{q^2+a}\right),
\end{equation}
which enables one to reduce the integrals to a form where, in the
three couples $\{i,l\}$, $\{j,m\}$ $\{k,n\}$, at least one element is
zero.

The previous procedure reduces the problem to the case $r=s=t=0$.
In this case, the integral is finite (and therefore of no interest for
us since we work in the minimal subtraction scheme) whenever the four
conditions $i+j+k+l+m+n>4$, $i+j+l+m>2$, $j+k+m+n>2$, $k+i+n+l>2$ are
satisfied simultaneously. The divergent integrals can be of two types:
\begin{enumerate}
\item 
if the smallest value between $i+l$, $j+m$, $k+n$ is zero, then the
integral factorizes, and identifies to a product of 1-loop integrals
of the form:
\begin{equation}
  \label{eq_int_1l}
  I_{i}(a)=\int_{\vect q}\frac 1 {(q^2+a)^i}=\frac 1{(4\pi)^{d/2}} 
\frac{\Gamma(i-d/2)}{\Gamma(i)}a^{d/2-i}
\end{equation}
\item if the smallest value between $i+l$, $j+m$, $k+n$ is 1, then the
  second smallest one must also be 1 for the integral to be divergent, and
  the last one is free. All these integrals can be calculated by using
  derivatives of the relation:
  \begin{equation}
    \begin{split}
      \int _{\vect q_1 \vect q_2 \vect q_3}\frac{\delta (\vect q_1
        +\vect q_2 +\vect q_3)}{(q_1^2 +a)(q_2^2 +b)(q_3^2 +c)}&=\\
      -C\frac{a^{d-3}+b^{d-3}+c^{d-3}}{d-3}\Bigg(\frac 1
        {2\epsilon^2}&-\frac 1 {4\epsilon}\Bigg)+\mathcal O(\epsilon^0)
    \end{split} 
\end{equation} 
with $C^{-1}=8\pi^2$.

\end{enumerate}

\subsection{Integrals for the 1-replica diagrams}

The situation for the 1-replica diagrams is in a sense simpler because
only one mass can appear. However, the vertices and propagators can now
come with the external momentum $\vect p$.  The first step in the
evaluation of the integrals consists in expanding the integrand in
powers of $\vect p$, keeping only the constant, linear and quadratic
terms which are of interest in the renormalization procedure. The
linear part of the integral vanishes for symmetry reasons. The
constant part can be evaluated by using the procedure described in the
previous subsection.

The quadratic part can come in the two following forms: $p^2$ or $(\vect
q_i\cdot \vect p)(\vect q_j\cdot \vect p)$. The first case is simple
and can be evalated by using the procedure described above. For the
second form, we used the relation
\begin{equation}
  \begin{split}
    \int_{\vect q_1\vect q_2\vect q_3}\delta(\vect q_1&+\vect
    q_2+\vect q_3)(\vect q_i\cdot \vect p)(\vect q_j\cdot \vect p)f =\\ 
    &\frac {p^2}d \int_{\vect q_1\vect q_2\vect q_3}\delta(\vect
    q_1+\vect q_2+\vect q_3)(\vect q_i\cdot\vect q_j)\ f
      \end{split}
\end{equation}
where $f$ is a function of $\vect q_1$, $\vect q_2$ and $\vect q_3$
and $i,j$ are equal to $1$, $2$ or $3$. The case $i\neq j$ can then be
treated as in the previous section. In the case $i=j$, the integrand
always comes with $(q_i^2+c)^\alpha$ in the denominator, so one can
use again the relation:
\begin{equation}
  \frac{q_i^2}{(q_i^2+c)^\alpha}=\frac 1{(q_i^2+c)^{\alpha-1}}-\frac
  c{(q_i^2+c)^\alpha} 
\end{equation}
and compute the remaining part by using the procedure described
previously.

\end{document}